\documentclass[fleqn,12pt]{wlscirep}
\usepackage[utf8]{inputenc}
\usepackage[T1]{fontenc}
\usepackage{setspace}
\usepackage{comment}
\usepackage{longtable}

\usepackage[capitalize]{cleveref} 

\definecolor{gray50}{RGB}{127, 127, 127}
\colorlet{gray50}{gray50}

\definecolor{purple}{RGB}{181, 23, 158}
\colorlet{purple}{purple}

\definecolor{blue}{RGB}{72, 149, 239}
\colorlet{blue}{blue}

\usepackage{xr} 

\usepackage[font=small,labelfont=bf]{caption}

\makeatletter
\newcommand*{\addFileDependency}[1]{
  \typeout{(#1)}
  \@addtofilelist{#1}
  \IfFileExists{#1}{}{\typeout{No file #1.}}
}
\makeatother

\newcommand*{\myexternaldocument}[1]{%
    \externaldocument{#1}%
    \addFileDependency{#1.tex}%
    \addFileDependency{#1.aux}%
}

\myexternaldocument{SI}

\title{The Clinical Trials Puzzle: How Network Effects Limit Drug Discovery}

\author[1]{Kishore Vasan}
\author[1, 2]{Deisy Morselli Gysi}
\author[1,2,3+]{Albert-László Barabási}
\affil[1]{Network Science Institute, Northeastern University, Boston, USA}
\affil[2]{Department of Medicine, Brigham and
Women’s Hospital, Harvard Medical School, Boston, United States}
\affil[3]{Department of Data and Network Science, Central European University, Hungary}
\affil[+]{a.barabasi@northeastern.edu}

\begin{abstract}

The depth of knowledge offered by post-genomic medicine has carried the promise of new drugs, and cures for multiple diseases. To explore the degree to which this capability has materialized, we extract meta-data from 356,403 clinical trials spanning four decades, aiming to offer mechanistic insights into the innovation practices in drug discovery. We find that convention dominates over innovation, as over 96\% of the recorded trials focus on previously tested drug targets, and the tested drugs target only 12\% of the human interactome. If current patterns persist, it would take 170 years to target all druggable proteins. We uncover two network-based fundamental mechanisms that currently limit target discovery: \textit{preferential attachment}, leading to the repeated exploration of previously targeted proteins; and \textit{local network effects}, limiting exploration to proteins interacting with highly explored proteins. We build on these insights to develop a quantitative network-based model of drug discovery. We demonstrate that the model is able to accurately recreate the exploration patterns observed in clinical trials. Most importantly, we show that a network-based search strategy can widen the scope of drug discovery by guiding exploration to novel proteins that are part of under explored regions in the human interactome. 

\end{abstract}

\begin{document}

\flushbottom
\maketitle
%
%
\thispagestyle{empty}

\section*{Introduction}

Prior to receiving approval by the Food and Drug Administration (FDA), a new drug must complete multiple phases of clinical trials to prove its efficacy and safety. 
The complete clinical trials pipeline for a single drug, from early safety testing to trials on large populations, takes on average six years\cite{dimasi2016innovation}, and is estimated to cost about \$1 billion USD\cite{khanna2012drug}.
In 2007, the FDA Act\cite{food2017food} required funders to publicly post clinical trial designs and results to an online repository managed by the National Library of Medicine (NLM), increasing transparency in the drug discovery process\cite{avorn2018fda}. Despite well-documented compliance issues on reporting the results\cite{weiland2020, casassus2021european, kozlovnih}, the accumulated data offers a unique lens into the drug innovation practices\cite{zarin2011clinicaltrials}, and has allowed researchers to conduct meta-analyses on disease specific trials\cite{cihoric2015hyperthermia, hirsch2013characteristics}, obtain key insights into equity for patients with rare diseases\cite{pasquali2012status, bell2014comparison}, and unveil systemic biases in patient demographics\cite{brady2020lack,kong2020gender}.

The choices in clinical trials, from designing the trial protocol to selecting the patient population to testing drugs for specific diseases, have direct implications for the efficacy and equity of drugs that enter the market. 
While advances in genomics, machine learning\cite{cao2013genome, jacob2008protein}, network medicine\cite{sonawane2019network, loscalzo2017network}, and pharmacology\cite{hopkins2008network} present novel opportunities for drug discovery, potentially reducing the cost and time of conducting exhaustive experimental testing\cite{chong2007new}, they may be inadequate if the discovered knowledge about drug candidates (\textit{in-silico}) is not actively transferred to applied settings (\textit{in-vitro}), and make their way into clinical practice. Therefore, understanding the drug exploration patterns documented by clinical trials is important to improve population health\cite{yao2010novel, vasan2021hidden}. 


In this work, we offer a large-scale temporal analysis of drugs and its target's trajectory through clinical trials by exploring the cumulative knowledge of the clinical trials database. 
By combining data from various sources, including investigational and approved drugs, rare and common diseases, proteins and its disease associations, we aim to understand the factors driving the discovery and exploration of new drugs and targets. We find that while the number of clinical trials continues to increase, the rate of novel drugs entering clinical trials has decreased since 2001, a puzzling effect potentially indicating a drug discovery winter. We also find that target selection is primarily driven by two distinct network-based mechanisms, preferential attachment and local network effects, leading to the over exploration of certain drugs and protein targets. Our results illustrate that we currently fail to utilize the complete therapeutic potential of the human genome, prompting us to offer a data-driven pathway to unlock its potential through the human interactome, which captures the physical interaction between targets. We build a quantitative model of drug discovery that helps unveil network effects capable of boosting the identification of novel targets.

\section*{Results}

\textbf{Curating clinical trials and drugs}

\begin{figure}[ht]
\centering
\includegraphics[width=\linewidth,height=42em,keepaspectratio]{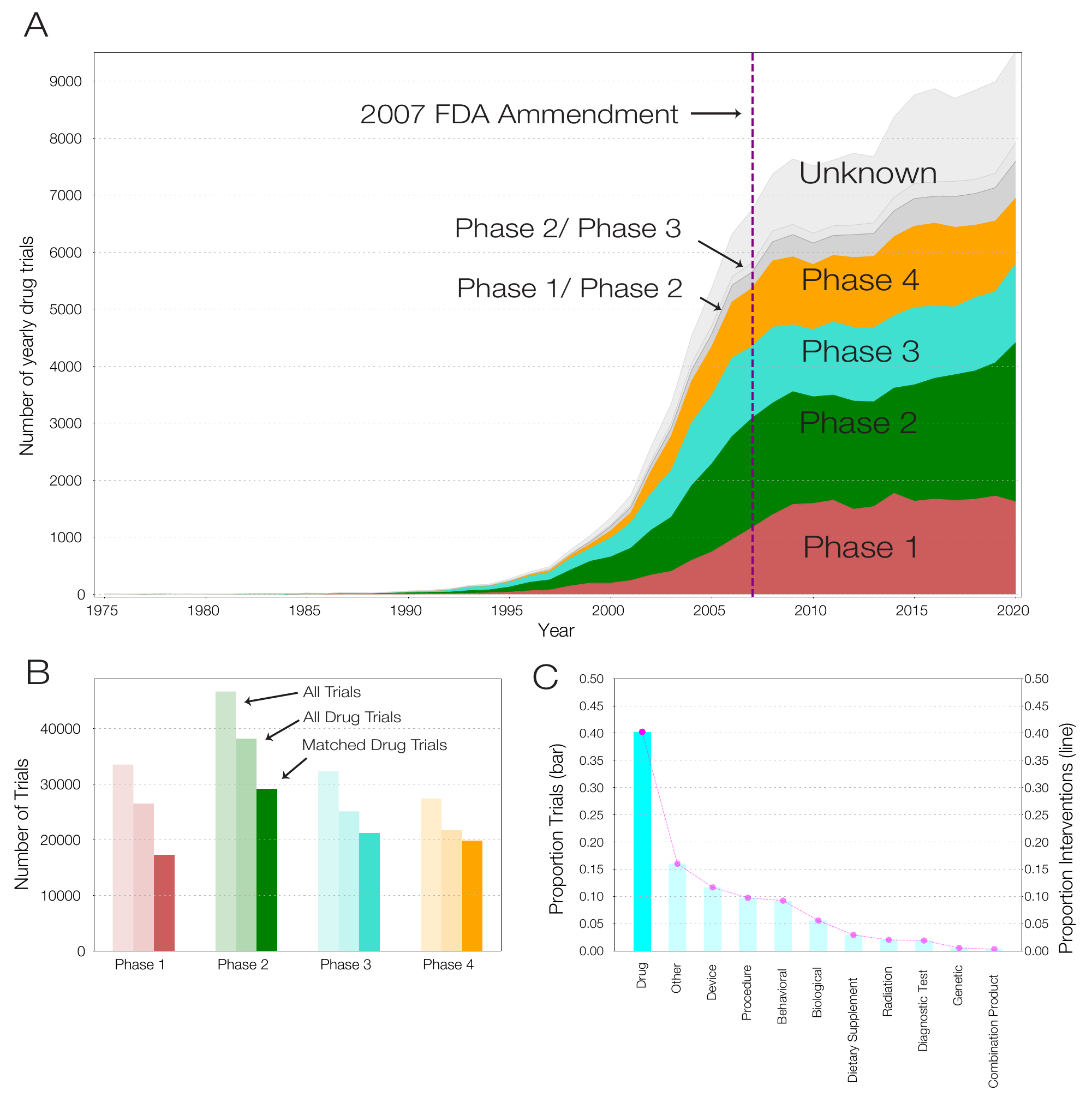}
\caption{\textbf{Clinical trials over time.} \textbf{(A)} Number of drug trials initiated over time. 
The rapid rise in clinical trials prior to 2007 is likely due to the 2007 FDA act that required all ongoing clinical trials to be registered on \url{clinicaltrials.gov} (purple line). 
We limit our analysis to phases 1 to 4 of clinical trials, and disregard combined phases and trials with unknown phase(gray). \textbf{(B)} Number of trials grouped by phase. We filter all known drug trials and match the drug interventions listed on the trials to known drugs (Supplementary Section 1.3). We show the final number of trials, grouped by phase, representing the corpus for our analysis (dark shade). \textbf{(C)} Proportion of trials and interventions by intervention type. Here we focus on drug trials, which represent roughly 40\% of all clinical trials and 30\% of all interventions.}
\label{fig:ct_counts}
\end{figure}

We extracted the clinical trials data from the publicly available clinical trials portal (\url{https://clinicaltrials.gov}), documenting 356,403 trials from 1975 to 2020. We observe a rapid growth in the number of reported drug trials before the 2007 activation date of the FDA amendment that required all funders to publicly disclose all active clinical trials by that year (Fig \ref{fig:ct_counts} A, vertical line), likely reflecting the sudden registration of all ongoing trials.
Following 2007, an organic growth sets in, indicating compliance with public reporting of new trials. 

We conducted a multi-step data standardization process to disambiguate drug names listed on trials (Supplementary Section 1), enabling the identification of 5,694 drugs used in 127,432 trials (89\% of drug trials). A drug is designed to bind to specific proteins in the human interactome, known as primary drug targets, responsible for the desired therapeutic effect. In some cases, drugs can also indirectly bind to other proteins, referred to as secondary drug targets. Of the 5,694 identified drugs, 2,528 (44\%) drugs have associations to 2,726 drug targets (both primary and secondary) and 1,442 (25\%) drugs have associations to 1,842 primary targets. We consider both primary and secondary targets, but we find that our results apply even when we limit our focus on primary targets only (Supplementary Section 1.2). 

Clinical trials are divided into several phases\cite{fdaclinical}. 
The pre-clinical stage (Phase 0 or early Phase 1) involves small dosage of a drug on a few people for a short duration to measure treatment response, corresponding to 1,880 (1.5\%) trials in our data. Phase 1 is the first full-scale human trial that includes close monitoring of treatment on a small number of patients, representing 26,207 trials (18\%). 
Phase 2 requires 25 to 100 patients with a specific disease condition to test for drug efficacy, representing 37,784 (26\%) trials. 
Phase 3 usually involves several hundred patients, where the experimental drugs are tested alongside other drugs to compare side effects and drug efficacy, representing 24,896 (17\%) trials. 
Finally, Phase 4 often involves thousands of patients, aiming to gain additional knowledge on drug safety over time, interaction with various diseases, and consists of 21,632 (15\%) trials. 
Some trials combined multiple phases such as Phase 1/ Phase 2, Phase 2/ Phase 3, together representing 11,381 (8\%) trials in our database. 
Here, we focus only on drug trials in Phases 1 to 4, representing in total 110,519 (76\%) trials (Fig \ref{fig:ct_counts} B highlighted), and disregard 19,718 (13\%) trials without  phase information (Fig \ref{fig:ct_counts} A, gray). 
Clinical trials can test multiple types of interventions, from drugs to medical devices to behavioral studies.
Drugs, the most widely tested intervention, represent 40\% of all trials, followed by medical devices (10\%) and behavioral interventions (10\%) (Fig \ref{fig:ct_counts} C). 

\textbf{Drug discovery winter}

\begin{figure}[ht]
\centering
\includegraphics[width=\linewidth]{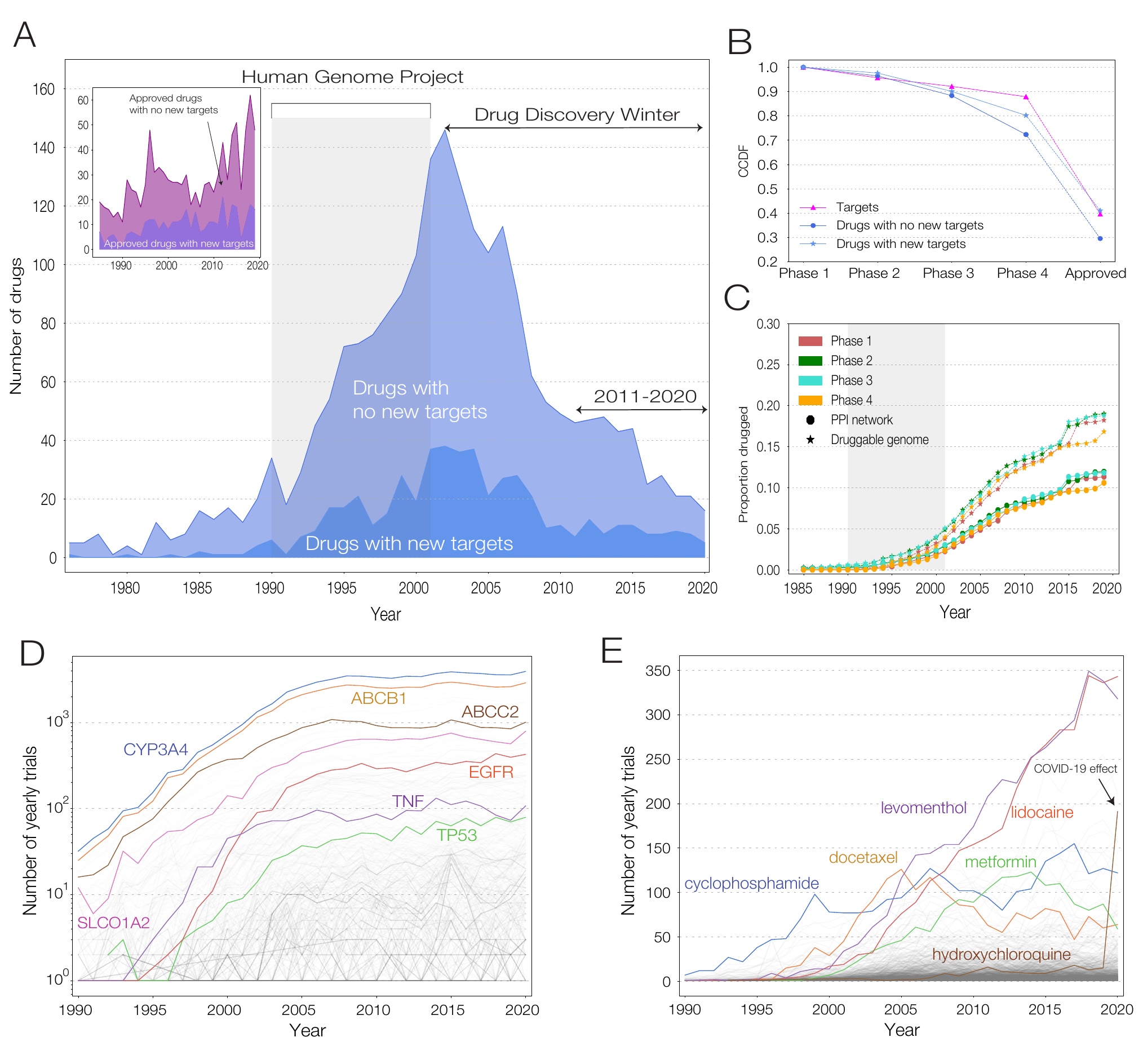}
\caption{\textbf{Drugs and targets tested in clinical trials.} \textbf{(A)} Number of drugs tested in clinical trials. We observe a slowdown in novel drugs tested since 2001, following the end of the Human Genome Project (HGP), signaling a drug discovery winter. For example, the number of drugs tested from 2011 to 2020 is considerably less compared to the exploration in previous decades. We also find an increasing gap between the number of approved drugs that have new targets and approved drugs with no new targets (inset).  
\textbf{(B)} Complementary cumulative distribution (CCDF) of the tested drugs and targets in each phase. We consider a protein and a drug in Phase 4 to have successfully completed Phase 1-3. The plot indicates that only a small proportion of drugs and targets in phase 4 have been approved. \textbf{(C)} Proportion of targets in the entire human genome in trials. We find that less than 20\% of all proteins have been tested in trials. The sudden jump in number of proteins in 2015 is due to a single publication in 2015 that found 306 targets for the drug \textit{fostamatinib} (see SI).
\textbf{(D)} Number of yearly trials of the top targets, demonstrating the inequality of drug exploration. Some targets, like CYP3A4, ABCB1, ABCC2 (highlighted) are the focus of multiple trials, while other targets are tested in only a few trials each year.
\textbf{(E)} Number of yearly trials for drugs. A select few drugs like levomenthol, lidocaine are tested in several trials every year, while other drugs are rarely tested. We see the impact of COVID-19 with a rapid increase in the number of trials for hydroxychloriquine.}
\label{fig:new_drugs_targets}
\end{figure}

The Human Genome Project (HGP), lasting from 1990 to 2001, boosted innovation and drug exploration\cite{gates2021wealth}, as in this decade clinical trials tested 768 (30\% of all) new drugs and 1,149 (42\% of all) new targets (Fig \ref{fig:new_drugs_targets} A shaded). Yet, beginning 2001, the exploration of new drugs has reduced. For example, between 2011 to 2020, clinical trials tested only 339 (13\%) new drugs and 662 (24\%) new targets (Fig \ref{fig:new_drugs_targets} A, Supplementary Fig \ref{fig:si_arima}), which, on average, corresponds to 33 new drugs and 24 new targets yearly, considerably lower compared to 99 drugs and 113 targets tested yearly in the early 2000s. Further, of the 339 new drugs that entered clinical trials, only 88 (25\%) drugs have novel targets, i.e. targeting not previously targeted proteins (Fig \ref{fig:new_drugs_targets} A, bottom). This indicates a drug discovery winter that started around 2001 characterized by a large number of clinical trials that focus mainly on drugs that target proteins already targeted by other previously tested or approved drugs.  

Throughout the history of clinical trials, 956 drugs (17\% of all), involving 1,340 targets (49\% of all) have been approved by the FDA (Fig \ref{fig:new_drugs_targets} A inset). 
Yet, only 342 (35\%) approved drugs test novel targets, indicating that drugs with established targets are more likely to receive approval\cite{krieger2018developing}. Although 1,449 (70\%) drugs and 2,076 (81\%) targets have reached Phase 4, only 40\% of those drugs and 51\% of those targets in Phase 4 targets received approval (Fig \ref{fig:new_drugs_targets} B). We also find that, on average a drug experiences a 3-year lag for approval after successfully completing Phase 3 clinical trials capturing the slow approval period, despite standard clinical development times\cite{brown2021clinical} (Supplementary Fig \ref{fig:si_time_to_approval_post_phase_3}). Taken together, we find that clinical trials have tested only 12\% of all human proteins and 22\% of all druggable proteins\cite{freshour2021integration} (Fig \ref{fig:new_drugs_targets} C). 
We estimate that if the current exploration patterns persist, it will likely lead to the exploration of 2,477 (13\% of all) proteins by 2025, and following this rate, it would take 170 years to test all 10,648 druggable proteins (Supplementary Section 2). 

\textbf{Previously tested proteins are repeatedly selected for future trials}

Clinical trials tend to focus on a small number of previously tested proteins, leading to an uneven approach to drug discovery (Fig \ref{fig:new_drugs_targets}, Supplementary Fig \ref{fig:si_gini_targets}, Supplementary Section 3). For example, we find that CYP3A4, ABCB1, ABCC2, SLCO1A2, proteins associated with the drug metabolism and transportation\cite{li2019current}, are involved in 72,884 (66\% of all) trials, while EGFR, TNF, TP53, proteins associated to auto-immune diseases and several neoplasms, are involved in 8,396 (8\% of all) trials (Fig \ref{fig:new_drugs_targets} D). Similarly, we find lidocaine, levomenthol, drugs that serve as anesthetics, to be over-represented in trials (Fig \ref{fig:new_drugs_targets} E). The COVID-19 pandemic had also a detectable impact on trial activity: hydroxychloroquine, a dormant drug which had a few clinical trials for over a decade, experienced a rapid increase in the number of trials in 2020\cite{thorlund2020real} (Fig \ref{fig:new_drugs_targets} E).

\begin{figure}[ht]
\centering
\includegraphics[width=\linewidth,
height=50em, keepaspectratio]{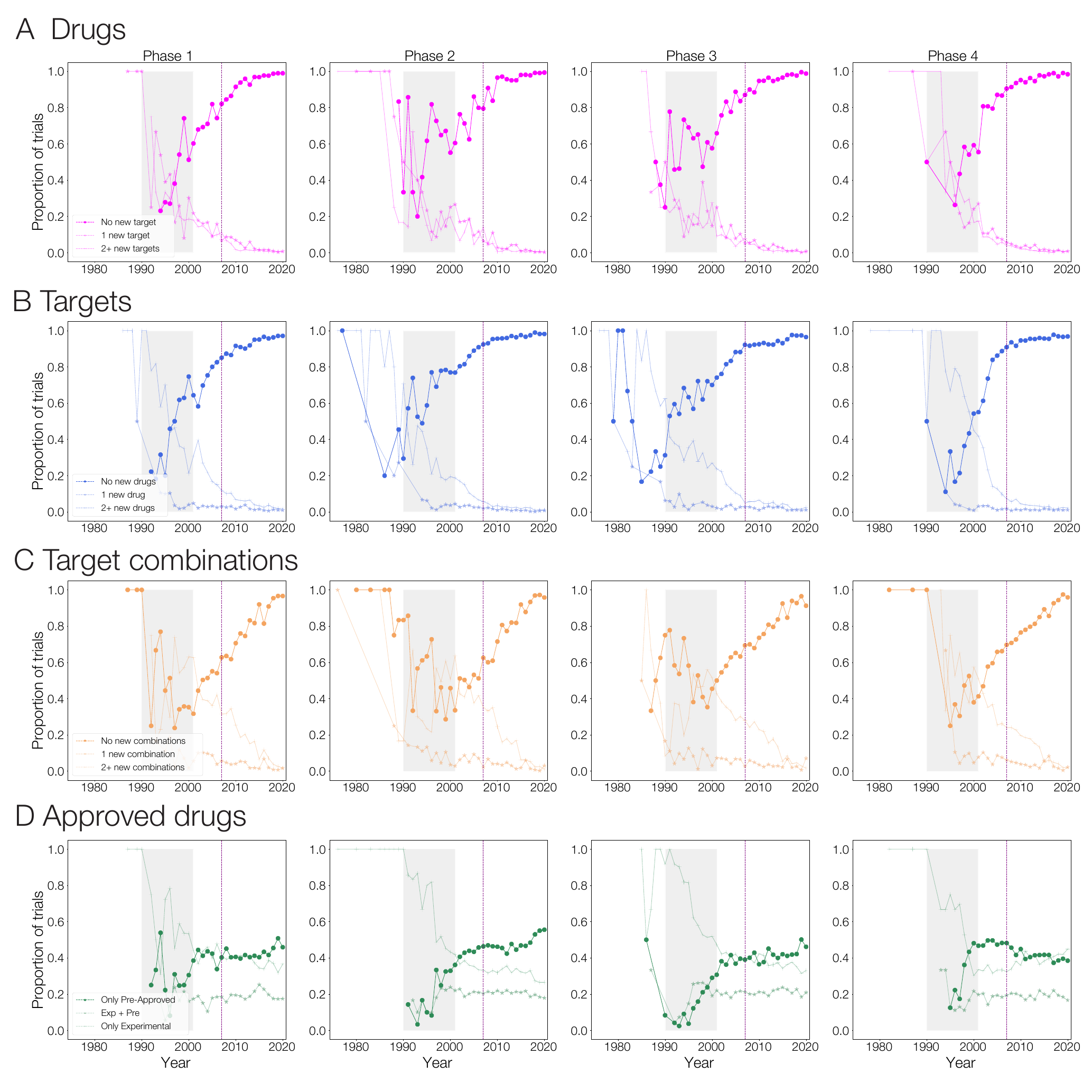}
\caption{\textbf{Novelty in clinical trials.} For each phase we identified the first time a drug, a target, or a target combination was first tested. We then trace the proportion of trials in each year for each phase that focus on \textbf{(A)} new targets \textbf{(B)} new drugs, and \textbf{(C)} new target combinations. Across (A-C), we observe a rapid rise in trials with new targets, coinciding with the completion of the HGP (shaded). However, beginning 2005, across all phases, a very small proportion of trials focus on novel targets, drugs, and combinations. \textbf{(D)} We also observe that close to half of the trials each year test previously approved drugs, indicating high interest in drug repurposing. This may partly be motivated by patent laws that force the patent owners to find new uses for the drug compound. As a consequence, we find growing inequality, where a select list of targets of approved drugs are repeatedly in clinical trials, thus preventing broad exploration of the human genome.}
\label{fig:prop_new_ct}
\end{figure}

A consequence of this uneven drug-target exploration is that only a small number of trials focus on new targets, new drugs, and new target combinations (Fig \ref{fig:prop_new_ct} A-C). The majority of the trials (50\%) involve only previously approved drugs, while 11\% of the trials test a combination of approved and experimental drugs (Fig \ref{fig:prop_new_ct} D; Supplementary Section 4). Seeking to find the patterns responsible for this over-exploration of previously targeted proteins, we measured to what degree targets that received more attention in the past are tested in subsequent years. We find that the number of drugs that target a specific protein, $N_{drug}(t)$, is well approximated by a growth rate following, $N_{drug}(t) \propto N_{drug}^{\gamma} (t-1)$, where $\gamma$ is a scaling exponent (Supplementary Fig \ref{fig:pa_evidence_targets}; $\gamma_{~2000} = 1.2$, $\gamma_{~2010} = 1.1$, $\gamma_{~2020} = 0.9$). This pattern, known as preferential attachment, is known to be responsible for the emergence of network hubs in network science\cite{merton1968matthew, barabasi1999emergence} and quantifies the degree to which previously tested proteins have a cumulative advantage over other proteins (Supplementary Section 5).


\textbf{The role of human interactome in drug exploration}

\begin{figure}[ht]
\centering
\includegraphics[width=36em, height = 39em, keepaspectratio]{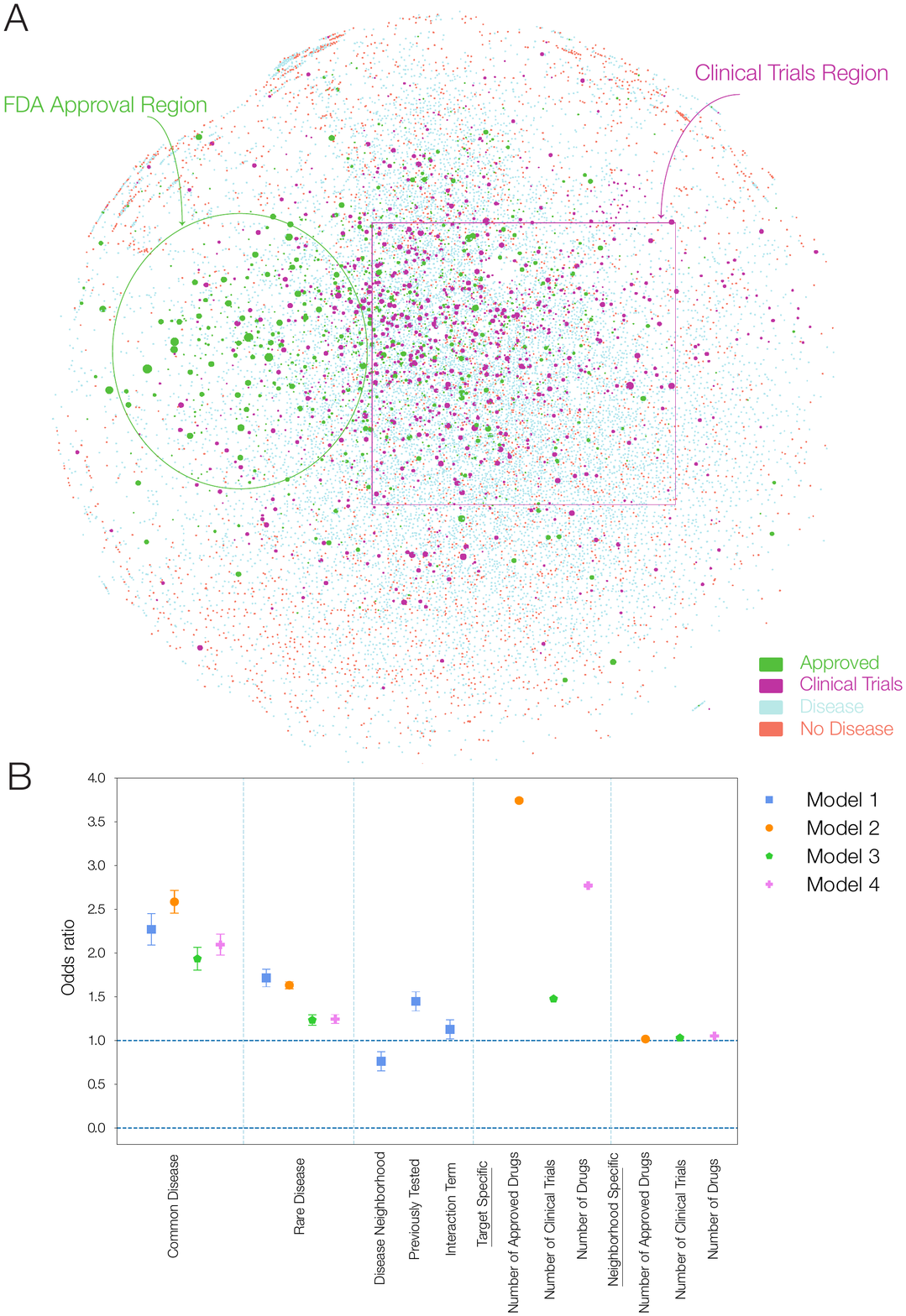}
\caption{\textbf{Networked exploration process of drug discovery. } \textbf{(A)} Protein-Protein Interaction (PPI) network. We observe that the region of proteins associated with FDA approved drugs (green) and proteins associated with experimental drugs (pink) are closely located in the network. We also find large unexplored regions (blue indicates disease associated proteins and purple indicates non-disease associated proteins). Nodes are sized based on number of clinical trials, indicating that the PPI network captures and potentially drives drug discovery. \textbf{(B)}. Logistic model results. We show the odds ratio estimate for different variables using four different models evaluating the likelihood of a protein to be selected for a new drug. Model 1 uses disease neighborhood variables: interactions to a disease associated protein and a previously targeted protein. Model 2 considers approval features: number of approved drugs that target the protein and its network neighborhood. Model 3 utilizes the clinical trials exploration: number of clinical trials of a protein and its network neighborhood. Model 4 uses drug exploration parameters: number of experimental drugs targeting the protein and its network neighborhood. The error bars indicate the standard error of the estimates. Results table shown at Table \ref{tab:reg_results}.}
\label{fig:ppi_net}
\end{figure}

Some diseases can be treated by inhibiting the disease associated proteins, but most often the effective drugs target proteins that are in the network vicinity of known disease proteins \cite{ay2007drug}. Indeed, most drugs act by perturbing the activity of the sub-cellular web known as the human interactome\cite{morselli2021network}, captured by experimentally detected Protein-Protein Interactions (PPI) (Fig \ref{fig:ppi_net} A). This prompts us to ask, can we take advantage of the interactome to better understand the patterns characterizing target discovery and exploration. To answer this question, we first mapped the 2,726 drug targets explored in clinical trials into the interactome, finding that 1,260 (92\% of all) experimental drugs target at least one protein that has been previously targeted by another approved drug, in line with Figs \ref{fig:new_drugs_targets} and \ref{fig:prop_new_ct}. However, when focusing on the proteins not targeted by previously approved drugs, we find that 891 (76\%) of them interact with at least one protein that is targeted by an approved drug, while 274 (23\%) are two steps away from the target of an approved drug. This local network-based clustering of experimental and approved drugs is absent if we randomly select the drug targets (Supplementary Section 6.1).  

We also find that proteins located farther from approved and experimental targets are rarely selected as a drug-target (Fig \ref{fig:ppi_net} A), even if they have multiple disease associations and are known to be druggable. In other words, we find a strong preference for targeting proteins that are embedded in local network neighborhoods with multiple explored targets (Supplementary Fig \ref{fig:si_prop_net_explored}). This means that a protein that interacts with other proteins that are the subject of multiple clinical trials for experimental or approved drugs is more likely to be selected as a new drug-target compared to a protein located in an unexplored network neighborhood. This suggests that the protein-protein interaction network captures and potentially drives drug discovery and exploration\cite{menche2015uncovering}. 


To unlock the impact of the observed network effects, we examine the likelihood of a protein to be selected as a drug-target in a future clinical trial using a Generalized Linear Mixed Model (GLMM). The GLMM model considers as input four features of each target: (1) disease associations, (2) number of approved drugs targeting it, (3) number of clinical trials it was involved in, and (4) number of experimental drugs targeting it. As an output, it offers several insights on the mechanisms governing new drug-target exploration (Fig \ref{fig:ppi_net} B; Supplementary Table \ref{tab:reg_results}):  

\begin{enumerate}
    \item Disease associated proteins are two times more likely to be in a clinical trial compared to proteins with no disease associations (OR: 2.2 [CI:1.6, 3.2], p<0.05). 
    \item Proteins experience increased likelihood of becoming the target of a new drug when they are already targeted by multiple approved drugs (OR: 3.7 [CI: 3.6, 3.9], p<0.01), multiple experimental drugs (OR: 2.7 [CI: 2.6, 2.8], p<0.01), or are the subject of multiple trials (OR: 1.47 [CI: 1.45, 1.49], p<0.01). 
    \item Previously untargeted proteins are more likely to be selected if they interact with proteins associated with multiple approved drugs (OR: 1.01 [CI: 0.99, 1.04]), multiple trials (OR: 1.03 [CI: 1.01, 1.04], p<0.01), or multiple experimental drugs (OR: 1.05 [CI: 1.03, 1.07], p<0.01). 
\end{enumerate}

These findings establish two fundamental mechanisms that drive drug exploration:

(i) \textit{Preferential attachment:} The future attractiveness of a protein as a drug candidate increases as more drugs target it and more trials focus on it (increased clinical exposure). For example, for a protein that is already targeted by ten drugs, its odds of being the target of a new drug increases eight-fold, compared to a protein not targeted by a drug. 

(ii) \textit{Local network effects}: Previously untargeted proteins located in network neighborhoods with high exploration patterns (containing multiple drug targets and clinical trials) are more likely to be selected as new drug target compared to proteins located in network neighborhoods with fewer clinical trials and drugs. 

\textbf{Modeling choices in drug discovery}

\begin{figure}[ht]
\centering
\includegraphics[width=\linewidth,height = 32em, keepaspectratio]{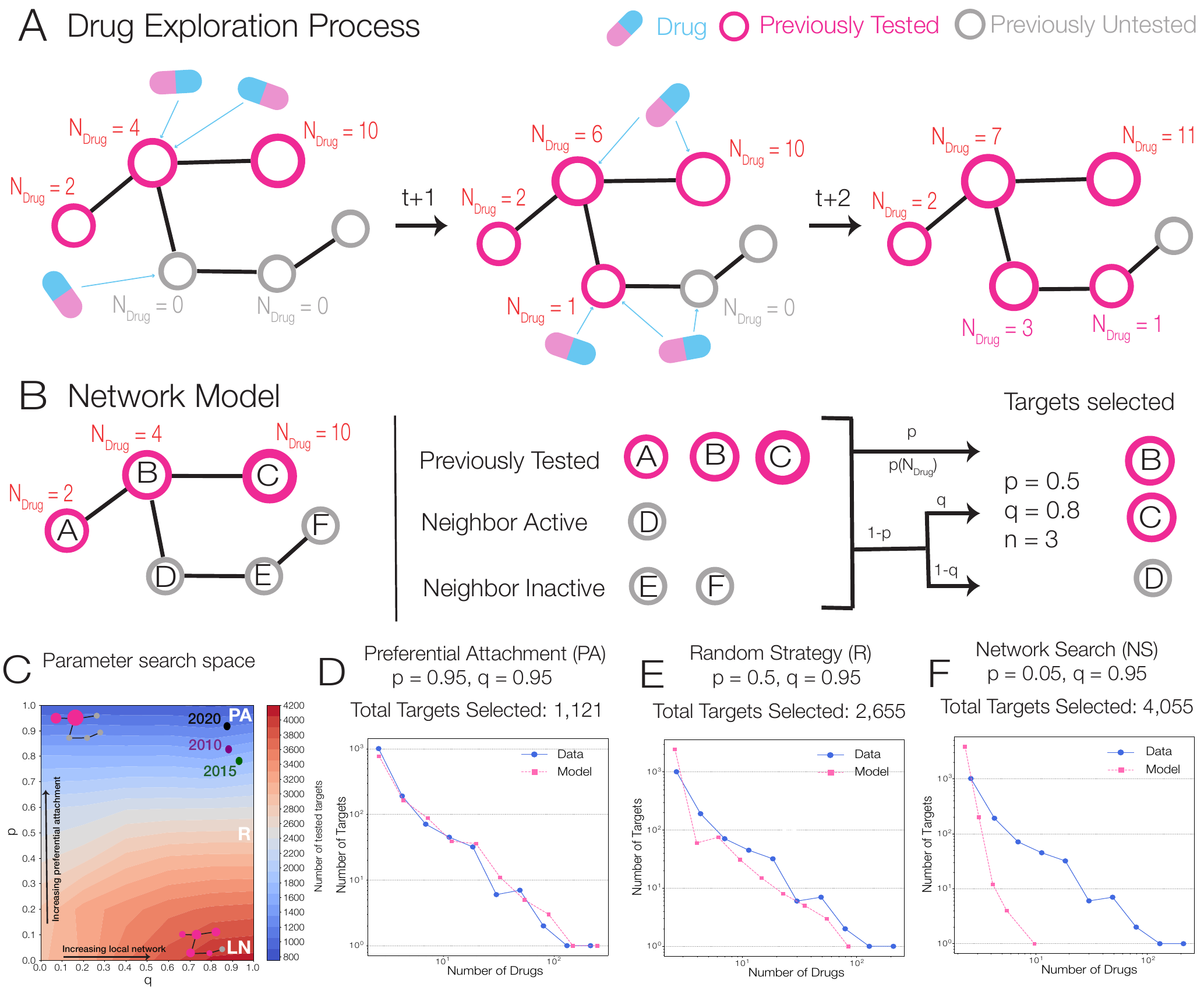}
\caption{\textbf{Modeling mechanisms of drug-target discovery.} \textbf{(A)} The exploration of the protein-protein interaction (PPI) network, where new proteins are selected as targets for drugs in clinical trials. For time $t_0$, we calculate the number of drugs that previously targeted each protein in the network, $N_{Drug} (t_0)$. At the timestep $t_0$, new drugs are introduced in clinical trials for testing. We identify the targets of these drugs at time of trial, represented using arrows, and update the number of drugs for proteins at the next time step, $N_{Drug} (t_1)$. Similarly, we identify the drugs introduced at time $t_1$ and its targets and update the number of drugs that target a protein at time $t_2$. The temporal characteristics of each protein allows us to capture the drug discovery process in clinical trials. \textbf{(B)}) Network model. We consider the network at time step, $t_0$, using the above described process and group proteins into three categories: (i) proteins that were previously tested (ii) proteins connected to a previously tested protein, and (iii) proteins that are not connected to a previously tested protein. With probability $p$, we select a previously tested protein, while with probability $q$ we select a protein connected to a previously tested protein, and with probability $1-q$, we select a protein not connected to a previously tested protein. When choosing a previously tested protein, we sample proteins proportional to the number of drugs that have previously targeted it, $P(N_{Drug})$, representing preferential attachment. In the network simulations, we select $m(t)$ proteins (calculated from data) and update the network at the end of each time step. We describe one version of the simulation where parameters $n=3,\; p = 0.5,\; q = 0.8$ are used to select the proteins $B$, $C$, and $D$ at next time step $t_1$. \textbf{(C)} The search space of exploration. We measure the number of targets that are tested in the simulations as a function of the parameters $p$ and $q$. The circles indicate the empirical choices for different years (2010, 2015, 2020). We show the distribution of number of drugs per target obtained under the three different exploration strategies: \textbf{(D)} Preferential attachment (PA) ($p=0.95,\; q=0.95$). \textbf{(E)} Random (R) ($p=0.5,\; q=0.95$) and  \textbf{(F)} Network Search (NS) ($p=0.05,\; q=0.95$).}

\label{fig:net_scenarios}
\end{figure}

We build on the insights (i) and (ii) to introduce a network model that aims to quantitatively recreate the observed patterns in drug exploration (Supplementary Section 7), and helps us understand how to accelerate drug discovery by exploring a wider set of druggable candidates. We begin by creating a timeline of drug discovery, accounting for the precise dates when targets became associated with drugs (Fig \ref{fig:net_scenarios} A). 
Using the proteins (nodes) and its interactions (links) in the PPI network as the underlying space of possible exploration, we model drug discovery through two parameters: The parameter $p$ represents the probability that a previously tested protein is selected again for clinical trials. Hence, for $p=0$, we model the scenario where we always choose untargeted proteins, while for $p=1$ we always select previously tested proteins as targets. The second parameter, $q$, represents the probability that we choose an untargeted protein that is part of an explored neighborhood, driven by local network search (Fig \ref{fig:net_scenarios} B). Hence, for $q=0$, we always select proteins from unexplored neighborhoods, while for $q=1$ we select proteins from previously explored neighborhoods. Finally, to account for preferential attachment in target selection, a previously tested protein is selected again as a target proportionally to the number of drugs that have targeted it in the previous years, $P(N_{drug}(t)) \propto P(N_{drug}(t-1))$ (Supplementary Fig \ref{fig:pa_evidence_targets}). 

The advantage of the proposed model is that we can explicitly extract the parameters $p$ and $q$ from the clinical trials data (Fig \ref{fig:net_scenarios} C). For example, in 2010, 295 proteins were tested in clinical trials, of which 244 (82\%) were tested in previous clinical trials, and we find that of the 51 previously untargeted proteins, 45 (88\%) interact with a previously tested protein, hence $p=0.82$ and $q=0.88$. We find that the empirically obtained ($p$,$q$) parameters are remarkably stable over time, indicating that previously tested proteins are in each year preferred at high rates ($p_{2010}=0.82$, $p_{2015}=0.78$, $p_{2020}=0.91$; Supplementary Fig \ref{fig:si_net_gene_exploration}). We also find that among the untargeted proteins, those interacting with other previously tested proteins are more likely to be selected ($q_{2010}=0.88$, $q_{2015} = 0.92$, $q_{2020} = 0.87$), allowing us to quantify the stable patterns characterizing drug discovery (Supplementary Fig \ref{fig:si_prop_net_explored}). As Fig \ref{fig:net_scenarios} C shows, the empirically observed patterns are stable in the high ($p$, $q$) regime, with a slight shift over time to higher values of $p$ and $q$, confirming an increasing trend to explore previously tested targets. 

We find that for the observed ($p^{*}$, $q^{*}$) values, the network model accurately reproduces the distribution of number of drugs per target (Fig \ref{fig:net_scenarios} D; KS-distance: 0.06; p<0.01). The model also allows us to test the relative importance of its building blocks. For example, if we remove the preferential selection of targets, the model fails to capture the drug exploration patterns (Supplementary Fig \ref{fig:si_pa_absence_model}), confirming that preferential attachment (PA) is a key ingredient of the current drug exploration strategy. The model also unveils the imperfections of the current target selection patterns: the PA strategy, which redirects attention and resources to previously tested proteins, only tests 21 new targets yearly on average. As a consequence, the same protein is explored as a target for a total of 175 (17\% of all) drugs ($GINI=0.65$), acting as a hub of drug discovery (Supplementary Section 7.1). Overall, the current strategy, by repeatedly targeting previously tested targets, fails to take advantage of the broader potential of the interactome to unveil potential novel targets. To validate the model, we quantified its ability to predict drug candidates for three autoimmune diseases - Rheumatoid Arthritis (RA), Crohn's Disease (CD), and Asthma (Supplementary Section 7.2). We find that the model accurately predicted novel candidates for these diseases with 70\% accuracy (Supplementary Fig \ref{fig:si_prec_cd_asthma}). Further, we validated the predicted proteins through an extensive literature search, finding them to be biologically relevant (Supplementary Table \ref{tab:novel_drug_candidates}). For example, the model identified protein \textit{NLRP3} as a potential drug candidate for RA, which has been shown to reduce RA-induced inflammation in animal models\cite{liu2020cinnamaldehyde}. These results demonstrate that a network strategy can be a useful mechanism to drive exploration towards proteins in druggable parts of the network.


Finally, we want to exploit the predictive power of the network model to explore how to incentivize a wider exploration of human interactome as potential targets. For this, we examine two alternative exploration strategies: (i) Random (R) strategy, when the newly tested proteins are randomly selected ($p=0.5$);  (ii) Network Search (NS) strategy, when untargeted proteins interacting with previously targeted proteins are preferred ($p=0.05$). In each case we keep $q=0.95$, as indicated by the empirical data. 

We find that the random (R) strategy selects more drug targets than currently tested (as captured by the PA strategy) (2,655 vs 1,121), offering an opportunity to deviate from the current distribution of number of drugs per target (Fig \ref{fig:net_scenarios} E, KS-distance: 0.22, p<0.01). Despite the randomness of the strategy, the same protein is selected as a target for 110 (11\% of all) drugs ($GINI=0.35$), indicating that the R strategy also focuses repeatedly on a few network hubs, a pattern similar to the one observed in the PA strategy (175). Overall, the R strategy tests more targets than PA but still results in an over-exploration of a few proteins, and hence offers minimal improvements compared to PA (Supplementary Fig \ref{fig:si_gini_model}). 

In contrast, we find that the network search (NS) strategy generates statistically different distribution of number of drugs per target (Fig \ref{fig:net_scenarios} F; KS-distance: 0.37; p<0.01). Most importantly, the strategy selected 4,055 targets, a three-fold increase in the number of selected targets compared to the PA strategy (1,121). Of those 4,055, we find that 3,922 (96\%) are new targets. Further, the NS strategy selects the same protein as a target for a maximum of 10 (1\% of all) drugs ($GINI:0.06$), significantly lower compared to the R (110) or PA (175) strategies. 




Overall, our results indicate that the current practice (PA) is inefficient in terms of exploring the human interactome, focusing most resources on a small number of highly explored protein targets. In contrast, a network search approach can improve the total number of tested targets by preventing the emergence of protein hubs in drug discovery and also attract attention to potential drug candidates, ultimately resulting in a wider exploration of the human interactome. These results suggest that policy changes, such as prioritizing the approval of drugs with novel targets or targeted funding from the National Institutes of Health (NIH) towards the exploration of novel targets, could significantly enhance drug discovery by re-focusing resources on a wider range of novel targets while maintaining accuracy.

\section*{Discussion}


A scientist's choice of an idea to pursue is influenced by a combination of the project novelty and its potential research impact\cite{rzhetsky2015choosing, fortunato2018science}.
Similarly, a pharmaceutical company's choice of a target for a new drug is influenced by its potential market value and the likelihood that the drug succeeds in clinical trials\cite{golec2007financial}. However, the high attrition rates of drugs in clinical trials\cite{kola2004can}, difficulties with patent licensing\cite{price2020cost}, and the growing cost of developing new molecules\cite{munos2009lessons} have led to a risk-averse approach to drug discovery characterized by `small bets, big wins'\cite{krieger2018developing}. While this strategy, resulting in the creation of multiple drugs within the same therapeutic class\cite{gagne2011many}, increases competition and reduces drug prices\cite{wertheimer2004pharmacoevolution, dimasi2011competitiveness}, it takes away resources from the exploration of novel drugs and targets\cite{naci2015drug}, encouraging incremental innovation and hindering progress for population health. 

Our analysis of clinical trials data shows that the highest growth in drug exploration was between 1990 and 2001, likely driven by the advent of the Human Genome Project (HGP). However, in the following two decades, there was a decrease in the incentive to test novel drugs, and a disproportionate focus on approved drugs (61\% of all trials). This allocation of resources ultimately slows the discovery of novel therapies. Further, drug discovery in clinical trials often prioritize previously tested proteins (preferential attachment) and proteins connected to previously tested proteins (network effect), neglecting proteins in under-explored regions of the network, even if they have disease associations and are verified as druggable targets. To optimize target exploration in druggable regions of the network and improve the number of tested targets, it may be beneficial to reduce the emphasis on previously tested proteins and adopt a network-based search for drug candidates.

Our proposed modeling approach offers a framework for economists, policymakers, and medical researchers seeking to optimize choices in drug discovery, particularly in situations with limited resources. The introduced drug discovery model could be extended to incorporate the exploration benefits of both successful and failed trials, results that are currently not systematically reported by pharmaceutical companies\cite{weiland2020, casassus2021european, kozlovnih}. Future work, ensuring data transparency, could incorporate multiple parameters on clinical trials, including de-identified information on trial participants to better inform drug discovery strategies. Optimizing the search strategy for drugs can help to maximize the potential of new drugs by targeting novel proteins within the human interactome.


\bibliography{biblio}

\section*{Author contributions statement}

All authors conceived and designed the experiments. K.V. conducted the data curation, analysis, and model simulations. All authors offered edits and approved the manuscript.

\section*{Data availability}

The data and code to replicate the analysis is available at \url{https://github.com/Barabasi-Lab/clinical\_trials}. 

\section*{Competing interests.}

A.-L.B. is the founder of Scipher Medicine and Foodome, companies that explore the use of network-based tools in health. K.V and D.M.G do not report any competing interests.

\flushbottom

\section*{\huge Supplementary Information}

\vspace{2em}
%
%
\thispagestyle{empty}

\setcounter{figure}{0}  
\setcounter{table}{0}  

\renewcommand{\thefigure}{S\arabic{figure}}

\renewcommand{\thetable}{S\arabic{table}}

\section{Data}

\subsection{Data collection and curation}

\textbf{Drugs and targets.} Data about experimental and validated drugs is provided by DrugBank\cite{wishart2018drugbank}. DrugBank is a web-enabled database containing comprehensive molecular information about drugs, their mechanisms, their interactions and their targets and is publicly accessible using an API key at (\url{www.drugbank.ca}). The January 2021 release of this database gives us a total list of 14,315 drugs, of which 7,755 drugs are associated with 4,265 targets.

Each drug-target map has a representative publication verifying the association, representing 51,839 publications. We then use the year of publication of each paper to recreate the temporal discovery process of each drug and its target associations. We combine this information with the drug trial year allowing us to accurately identify targets that were tested in each year.

\begin{figure}[ht]
\centering
\includegraphics[width=\linewidth, height = 25em, keepaspectratio]{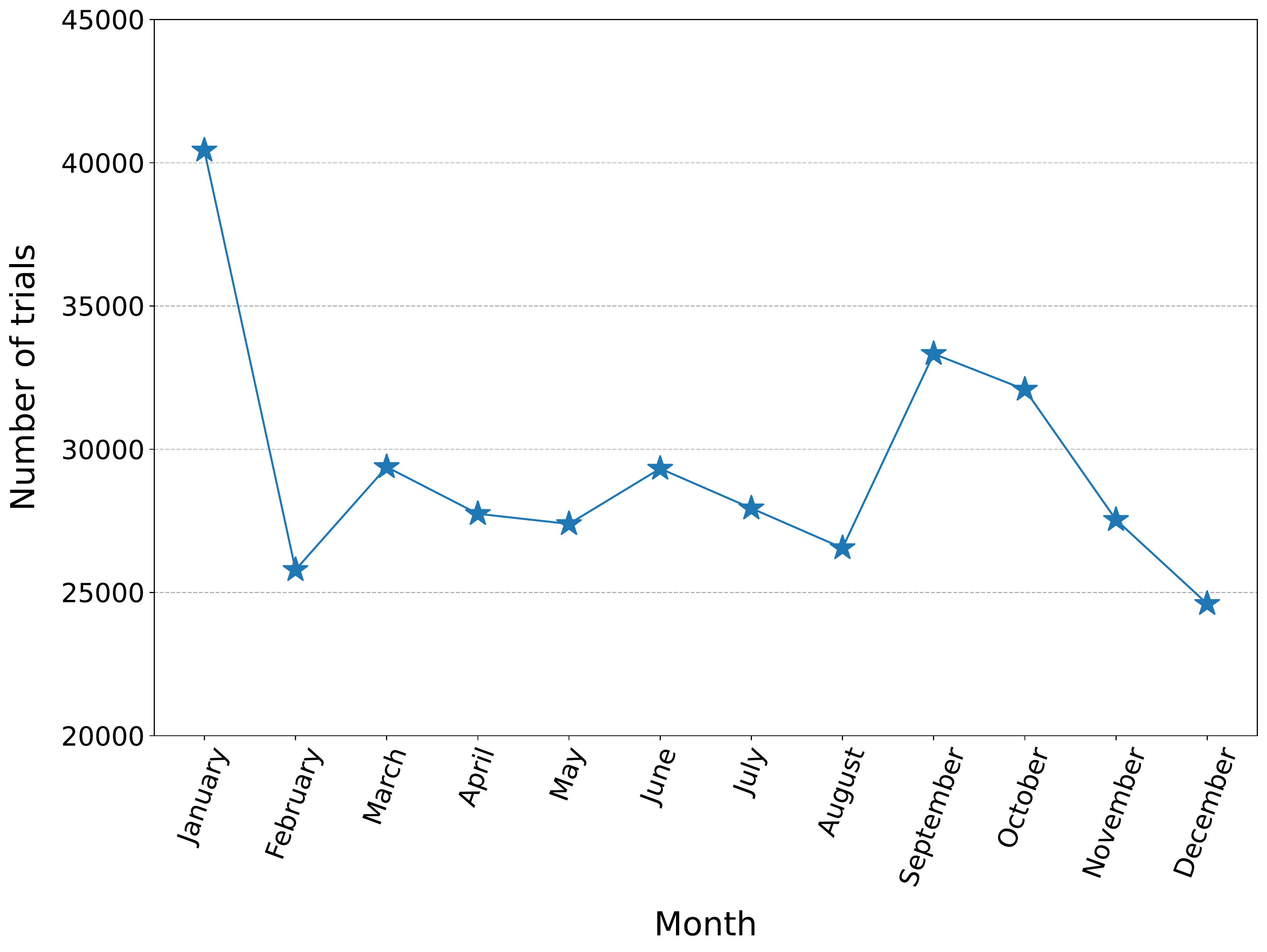}
\caption{\textbf{Number of trials grouped by month.} We find that majority of the trials started in January, a peak in September and a decline in December. }
\label{fig:trials_month}
\end{figure}

\begin{figure}[ht]
\centering
\includegraphics[width=\linewidth, height = 25em, keepaspectratio]{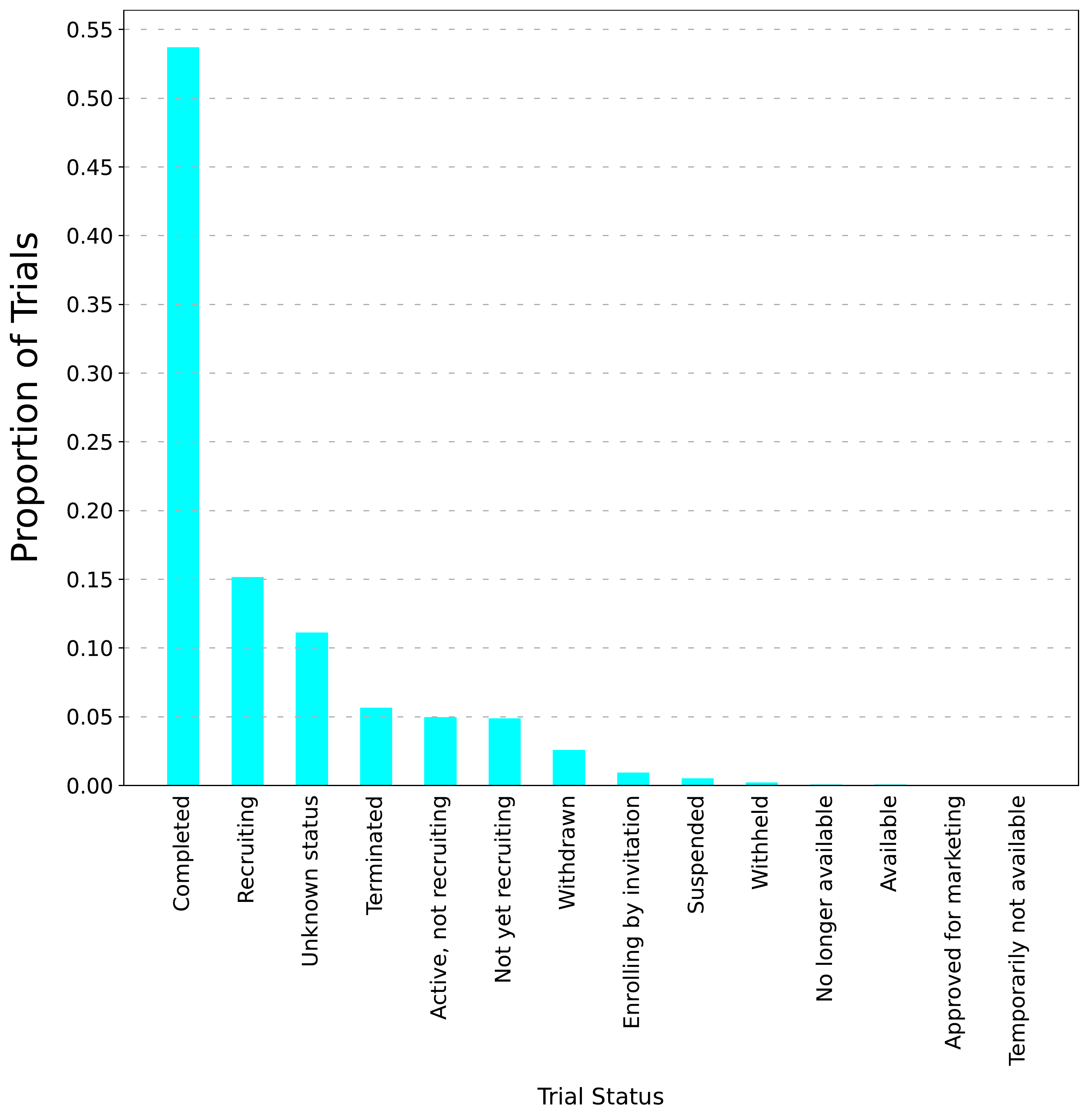}
\caption{\textbf{Status of trials.} We find that majority of the trials in our data are completed. A very small proportion of trials are terminated, suspended, or withdrawn. }
\label{fig:trials_state}
\end{figure}

\begin{figure}[ht]
\centering
\includegraphics[width=\linewidth, height = 45em, keepaspectratio]{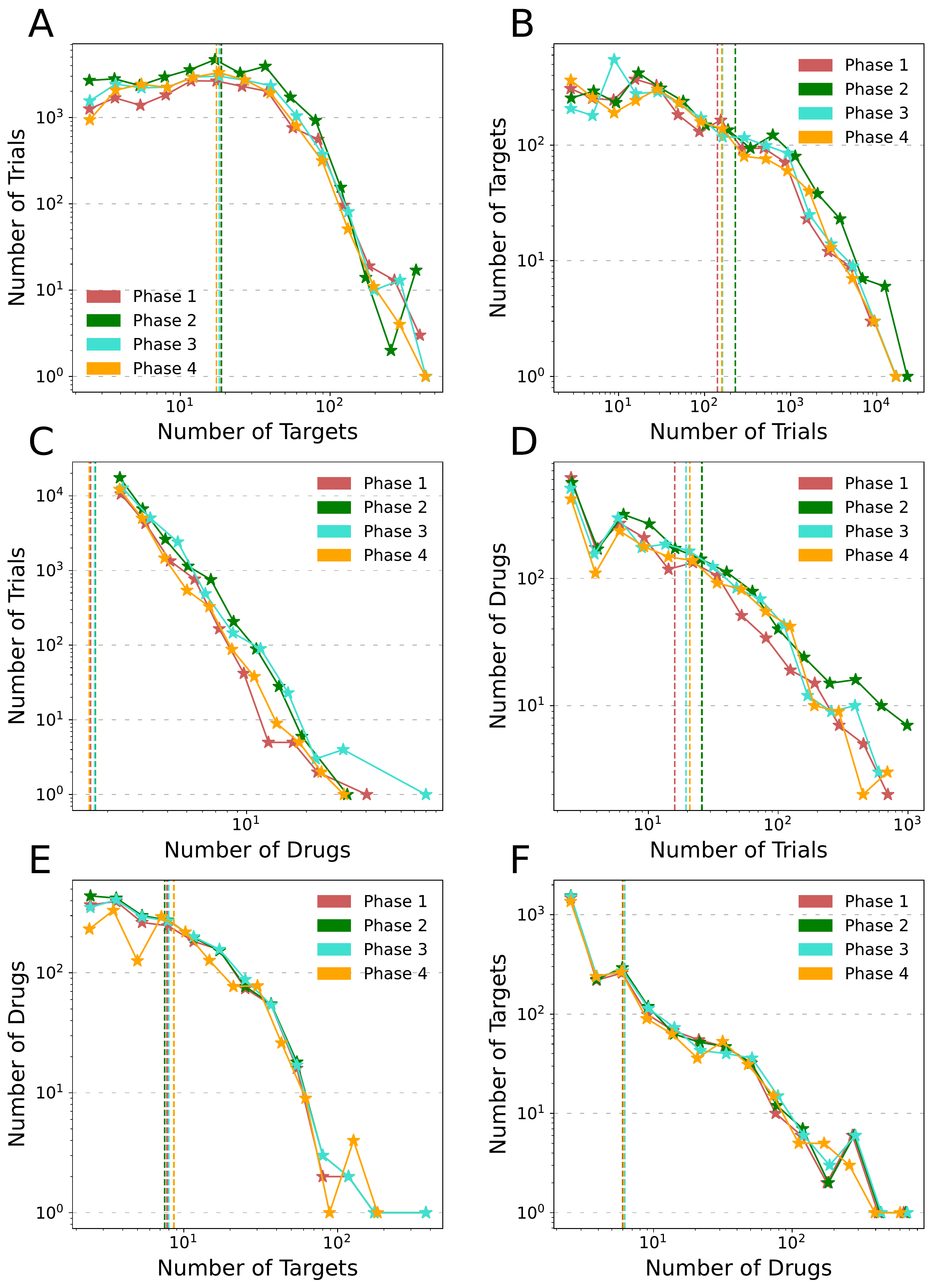}
\caption{\textbf{Distribution plots for different phases.} \textbf{(A)} Number of targets in number of trials. \textbf{(B)} Number of trials versus number of targets. \textbf{(C)} Number of drugs versus number of trials. \textbf{(D)} Number of trials versus number of drugs. \textbf{(E)} Number of targets versus number of drugs. \textbf{(F)} Number of drugs versus number of targets. The lines indicate the average of each group (Phase 1, Phase 2, Phase 3, Phase 4). }
\label{fig:si_distribution}
\end{figure}

\textbf{Clinical trials data.} The contents of all listed 356,403 clinical trials was downloaded on November 1, 2020 from (\url{https://clinicaltrials.gov}). All of the studies are grouped using NCT id which serves as the identifier for each trial (Supplementary Fig \ref{fig:si_distribution}). Every trial contains information about the date of trial (Supplementary Fig \ref{fig:trials_month}), type of trial (e.g. intervention, observational), its associated phase (e.g. Phase 1, Phase 2), status (e.g. completed, recruiting) (Supplementary Fig \ref{fig:trials_state}), a list of conditions (e.g. asthma, rheumatoid arthritis), a list of interventions (if applicable, e.g. budenoside, inhaler) and its associated types (e.g. drug, medical device). We then filter all trials that have a ``drug'' type associated with any of its listed interventions. This gives a subset of 146,314 trials. 

\textbf{Clinical trials drug data curation.} The listed drug names part of clinical trials are not standardized, and presents an issue to accurately identify drug exploration. For example, the drug `lepirudin' may be refered to as `lepirudin recombinant', `hirudin variant-1' or even its associated brand name 'Refludan'. As a result, we find a total list of 94,615 interventions in the clinical much higher than the number of drugs identified by DrugBank. To standardize the drug names, we conduct a multi-step matching process. First, we map the intervention names to the direct name on drug bank, giving us a total of 103,398 (70.6\% of all drug trials) trials and 4,458 drugs. Next, we map the intervention names to the drug synonyms provided by DrugBank allowing us to map an additional 7,698 trials. We also connect the drug names to the official drug product names allowing us to map another 14,759 trials. We also map intervention names with the wikipedia names of drugs providing additional drug maps for 500 drugs. Finally, we map the drugs names with a fuzzy match with drug names, providing mapping for another 1,077 trials. At the end of this methodology, we are left with 127,432 trials (87.6\% of all) and 5,694 drugs. We also control for placebo drugs in trials by searching for the term 'placebo' in the intervention names. We thus remove 1,171 trials on 590 drugs from our analysis. 

The data curation steps then reveal 127,432 drug trials for 5,694 drugs and 2,726 targets, representing the final data used in the analysis. The data is available at \url{https://github.com/Barabasi-Lab/clinical\_trials}, along with the code needed to recreate the analysis.

\textbf{Druggable genes.} The list of druggable genes is curated by a large-scale crowdsourcing effort by incorporating multiple data sources (e.g. Gene Ontology, OncoKB, PharmGKB)\cite{freshour2021integration}. The data is publicly available for free download from DG-IDB(\url{www.dgidb.org}) The November 2020 version of the data update was extracted for our analysis which contains 10,648 druggable human proteins. 
It is important to note that the finding of a drug-gene interaction as potentially druggable does not necessitate the ineffectiveness (or the lack thereof) for a drug to interact with other genes in different regions. 

\textbf{Protein-protein interaction network.} The proteins in the cell of an organism are known to have biological interactions with other proteins in neighboring cells. 
This relationship between proteins can be mapped to represent a network of genes and its interactions, a well-studied mechanism in network medicine\cite{gysi2020network}. The protein interaction network comprises of 18,508 nodes (proteins) and 332,646 edges (interactions). 

\textbf{Drug approval data.} The data regarding drugs and its approval is provided by the Food and Drug Administration (FDA).\footnote{the data is publicly available at \url{https://www.fda.gov/drugs/development-approval-process-drugs/drug-approvals-and-databases.}} The entire corpus was extracted in December 2020 that contains 1,002 approved drugs. Out of these drugs, we found 911 drugs in the clinical trials data. 

\begin{figure}[ht]
\centering
\includegraphics[width=30em, keepaspectratio]{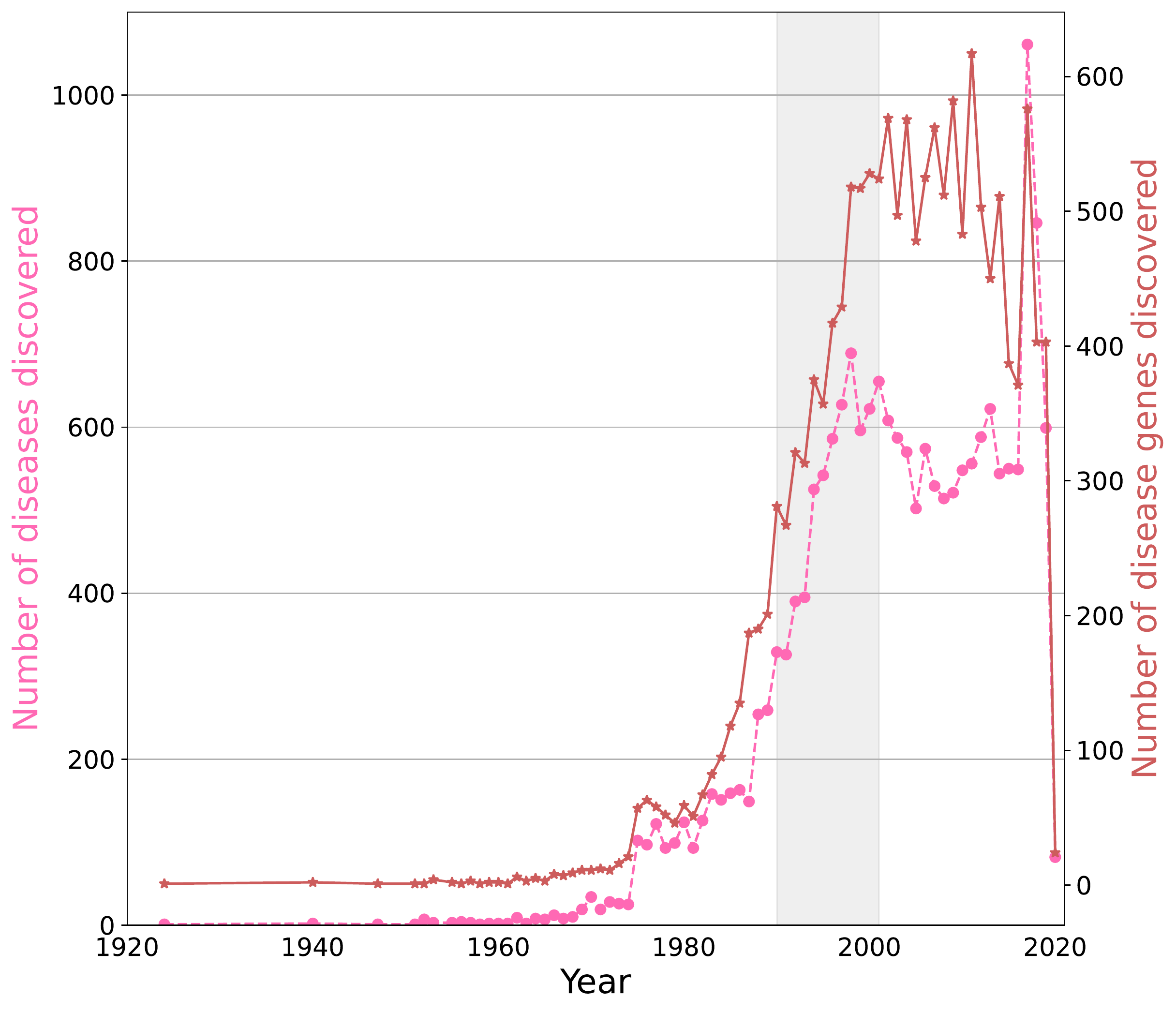}
\caption{\textbf{Temporal discovery of diseases and disease-associated genes.} We use the DisGeneNet data to identify the number of diseases discovered (left) and the number of gene-disease associations discovered (right) each year.}
\label{fig:si_disease_disc}
\end{figure}

\textbf{Disease data.} The data about disease associations were extracted from DisGeneNet\cite{pinero2020disgenet}. We find 15,474 genes associated with 19,620 diseases. Since the data also lists the corresponding publication reference that discovered the disease association, we map the publication (PubMed) id with the year of publication to identify the specific year that the gene was found to be associated with a disease, allowing us to accurately recreate the exploration patterns (Supplementary Fig \ref{fig:si_disease_disc}).

\begin{figure}[ht]
\centering
\includegraphics[width=30em, keepaspectratio]{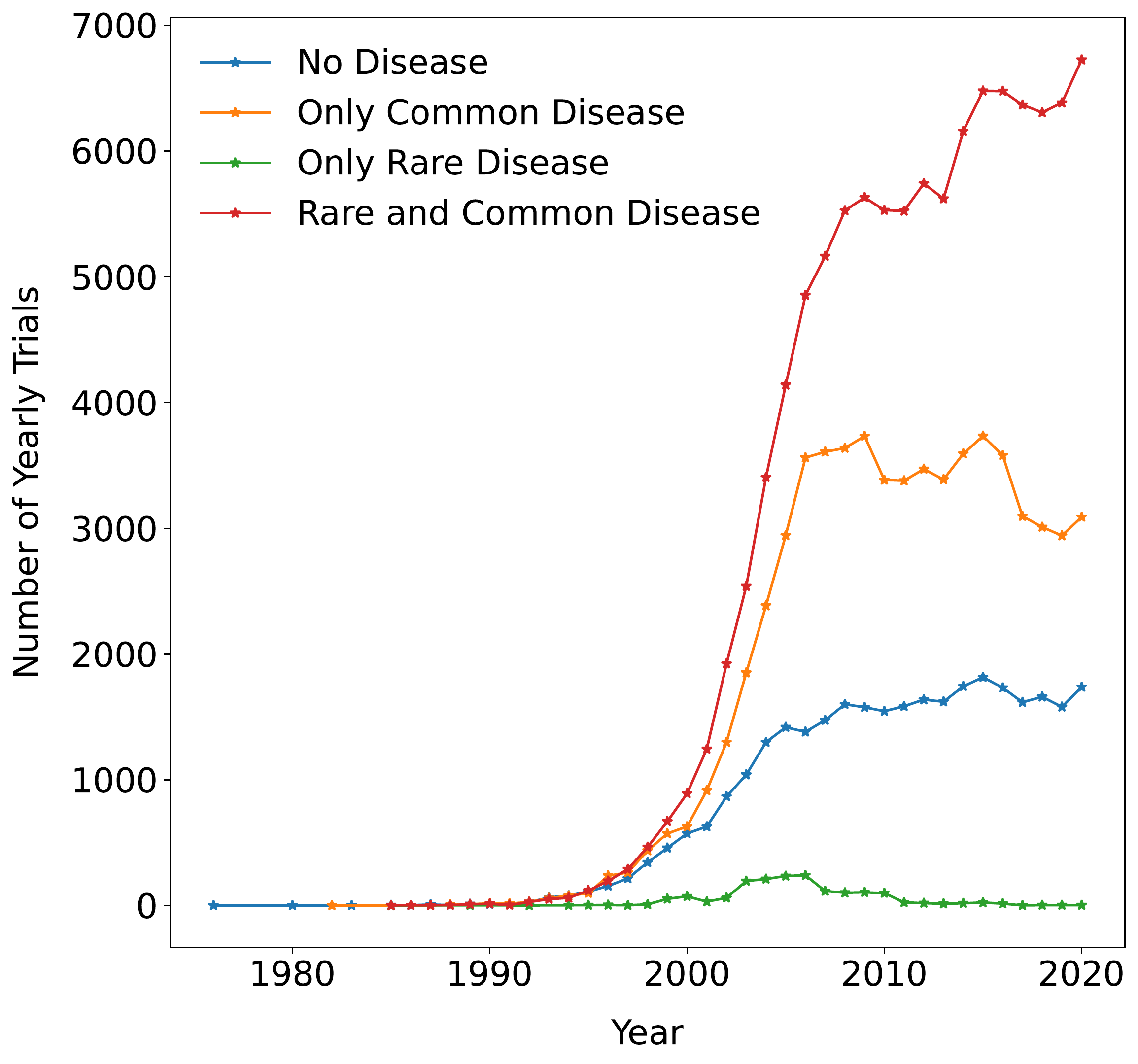}
\caption{\textbf{Number of trials for rare and common disease genes.} We observe that genes that are associated to only rare diseases have very few trials compared to genes associated with common diseases. }
\label{fig:rare_common_trials}
\end{figure}

\textbf{Common and rare diseases.} Information about common and rare diseases were extracted from \textit{Orphanet: an online rare disease and orphan drug data base.} The data is indexed via ORPHAcode that links diseases to associated genes, along with information about the association like causative, modifier, susceptibility. We then map these diseases with the DisGeneNet data through mesh ID to discover gene associations with rare and common diseases (Supplementary Fig \ref{fig:rare_common_trials}). After mapping, we find 29,001 common diseases associated with 15,339 genes and 1,169 rare diseases associated with 9,152 genes. The data is free to download from  \url{http://www.orpha.net}. Accessed on September 2021. 

\begin{figure}[ht]
\centering
\includegraphics[width=30em, keepaspectratio]{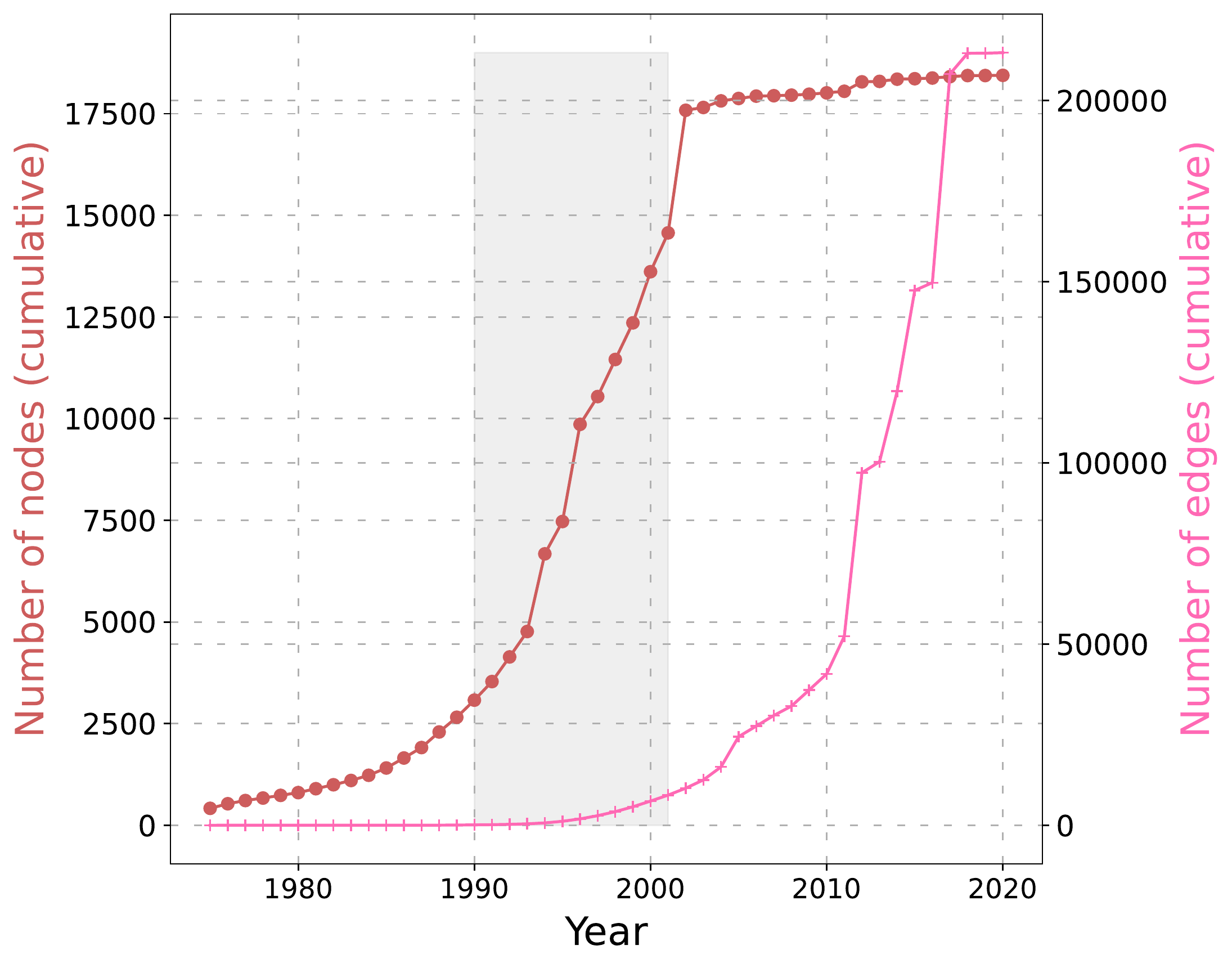}
\caption{\textbf{Temporal discovery of genes and its associations in the PPI network.} }
\label{fig:si_ppi_disc}
\end{figure}

\textbf{Timeline of protein discovery and interactions.} The data regarding the year of discovery of proteins and its interactions is collected by parsing 702,320 publications from the PubMed database\cite{gates2021wealth} (Supplementary Fig \ref{fig:si_ppi_disc}). This data allows us to recreate the temporal PPI network, accounting for the precise time a protein and its interactions were discovered. 

\subsection{Selecting primary targets}

Targets associated with drugs may have multiple mechanisms of action, such as, inhibitor, binder, activator, blocker, but for some drug-target associations the mechanisms of action may be unknown. The subset of targets with unknown interactions are referred to as 'off-target' genes and those with known interactions as 'primary-target' genes. In our data, we find that 98,139 (85\% of mapped) trials featuring 1,442 (57\% of drugs with targets) and 928 (34\% of all) primary targets. To consider only the subset of primary targets, we must disregard more than 40\% of the drugs in trials and 65\% of all targets, a large proportion of lost information.

\begin{figure}[ht]
\centering
\includegraphics[width=32em, keepaspectratio]{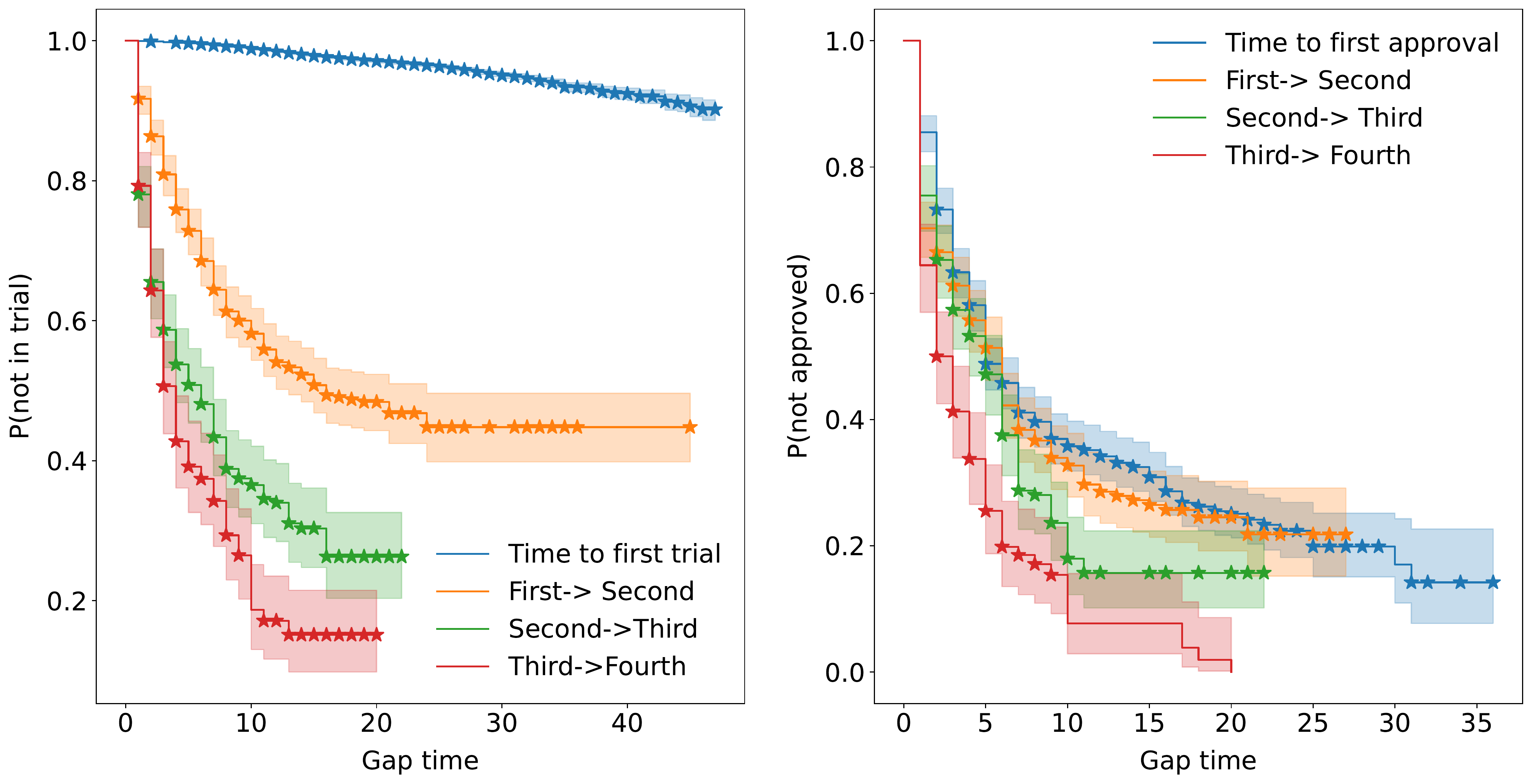}
\caption{\textbf{Repeated exploration of primary targets.} We consider all proteins that are experimentally verified as the primary targets of drugs. \textbf{(A)} Gap time in exploration, measured as the time spent in discovery period. \textbf{(B)} Gap time in approval, measured as the time spent in exploration period. This indicates that targets, primary or secondary, suffer from visibility bias, hence following the first experimental drug or first approved drug-targeting it, the gap time for subsequent drugs decreases. }
\label{fig:surv_plots_primary}
\end{figure}

Yet, considering both primary and off-target proteins for analysis may be important to map the space of drug exploration. For example, a drug capable of modifying the activity of an off-target may provide repurposing opportunities for that drug\cite{lounkine2012large}, and may initiate future exploration of that protein. Hence, we consider both primary and off-target proteins as part of the explored proteome. We provide the results in the supplementary text when we only consider primary targets, and find that the main findings do not change (Supplementary Supplementary Fig \ref{fig:surv_plots_primary}, Supplementary Table \ref{tab:reg_results_primary}). 

\subsection{Fostamatinib outlier}

\begin{figure}[ht]
\centering
\includegraphics[width=25em, keepaspectratio]{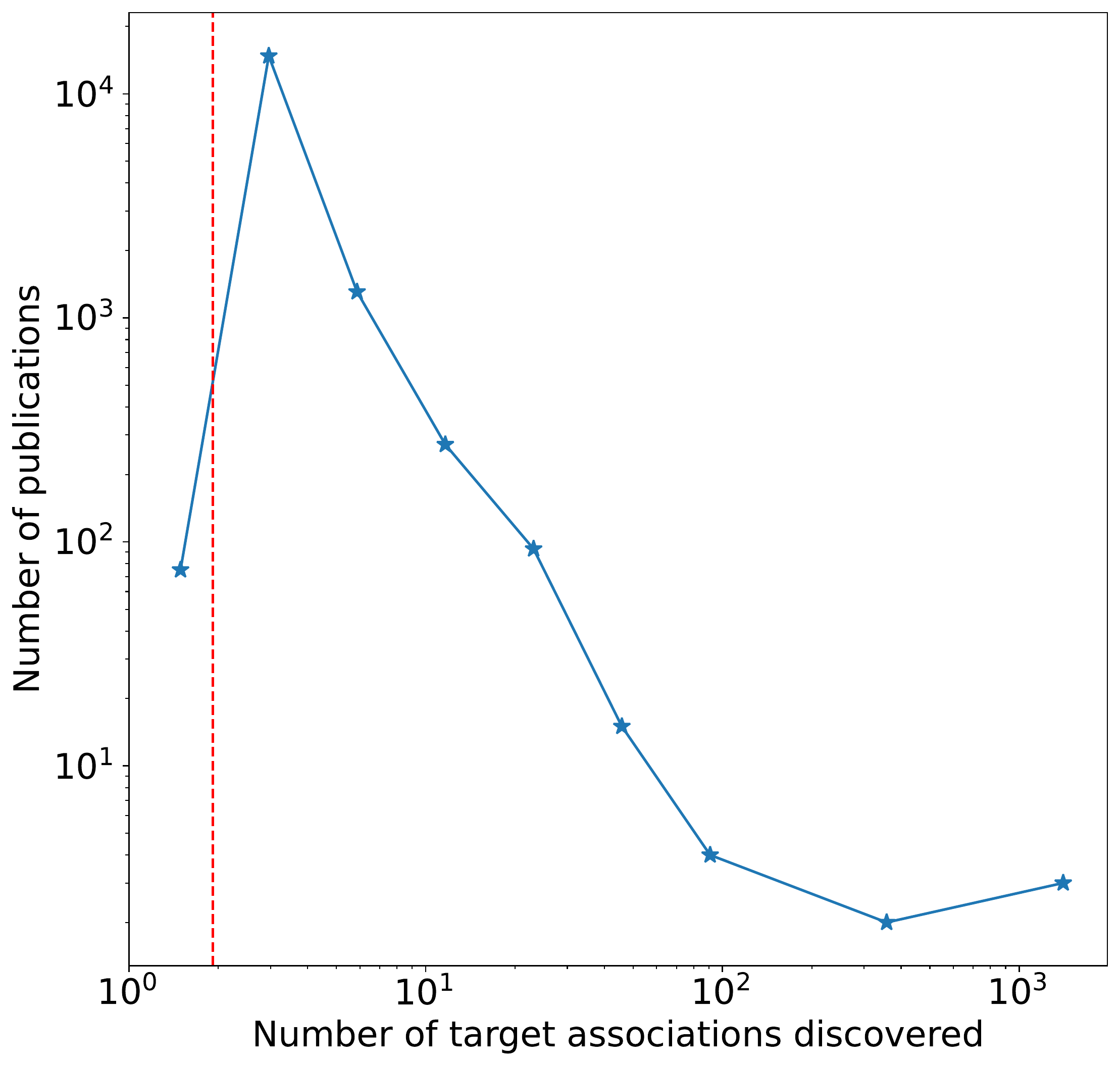}
\caption{\textbf{Distribution of target associations discovered in publications.} We find that on average every publication claims associations for 2 new targets for drugs (red line). Indeed, publishing associations for several hundred targets is highly unusual. }
\label{fig:si_pub_targets}
\end{figure}

We consider the year of drug-target association to build the target exploration in clinical trials. 
We do this byextracting metadata from PubMed of publications that provide verification for target associations for drugs. 
As we note in main text, clinical trials experience a sudden jump in exploration, attributed to the drug \textit{fostamatinib}, that was in trial in 2015 and found its approval in 2018. In 2015, the publication titled, "\textit{In vitro pharmacological profiling of R406 identifies molecular targets underlying the clinical effects of fostamatinib}"\cite{rolf2015vitro}, claimed 306 target associations for the drug \textit{fostamatinib}. Indeed, it is very unlikely for publications to claim associations for several hundred targets (Supplementary Supplementary Fig \ref{fig:si_pub_targets}). We remove this publication from our data in the subsequent analysis. 

\section{Estimating drug exploration}

\begin{figure}[ht]
\centering
\includegraphics[width=25em, keepaspectratio]{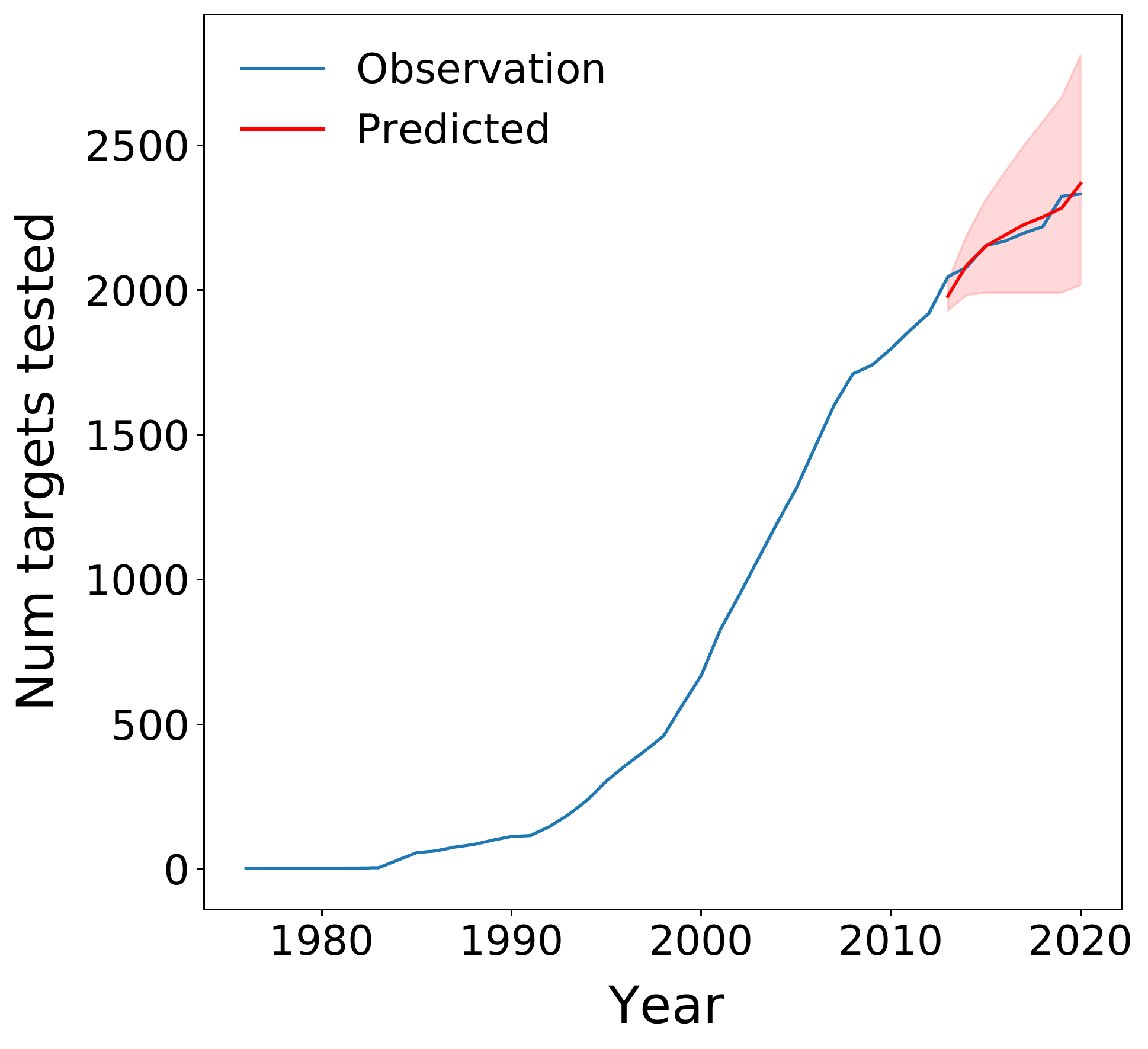}
\caption{\textbf{Time series prediction of number of targets tested.} We use a best fit ARIMA model to estimate the number of targets tested. The blue lines is the observed counts and the red line indicates the predicted values.}
\label{fig:si_arima}
\end{figure}

We utilize an Auto-Regressive Integrated Moving Average (ARIMA) time series model\cite{box2015time} to predict the drug exploration patterns. The model accounts for seasonal variation in trends to forecast future events. We consider the number of new targets tested every year as the output variable and estimate the best model fit using root mean square error estimation (RMSE). We utilize 80\% as training data and find the best model fit (1, 0, 3) with RMSE 50 (Supplementary Supplementary Fig \ref{fig:si_arima}). The model estimates that by 2025, 2,477 targets will be tested (95\% confidence interval: 2,445 - 3,682).

\section{Inequality in target and drug exploration}

\begin{figure}[ht]
\centering
\includegraphics[width=40em]{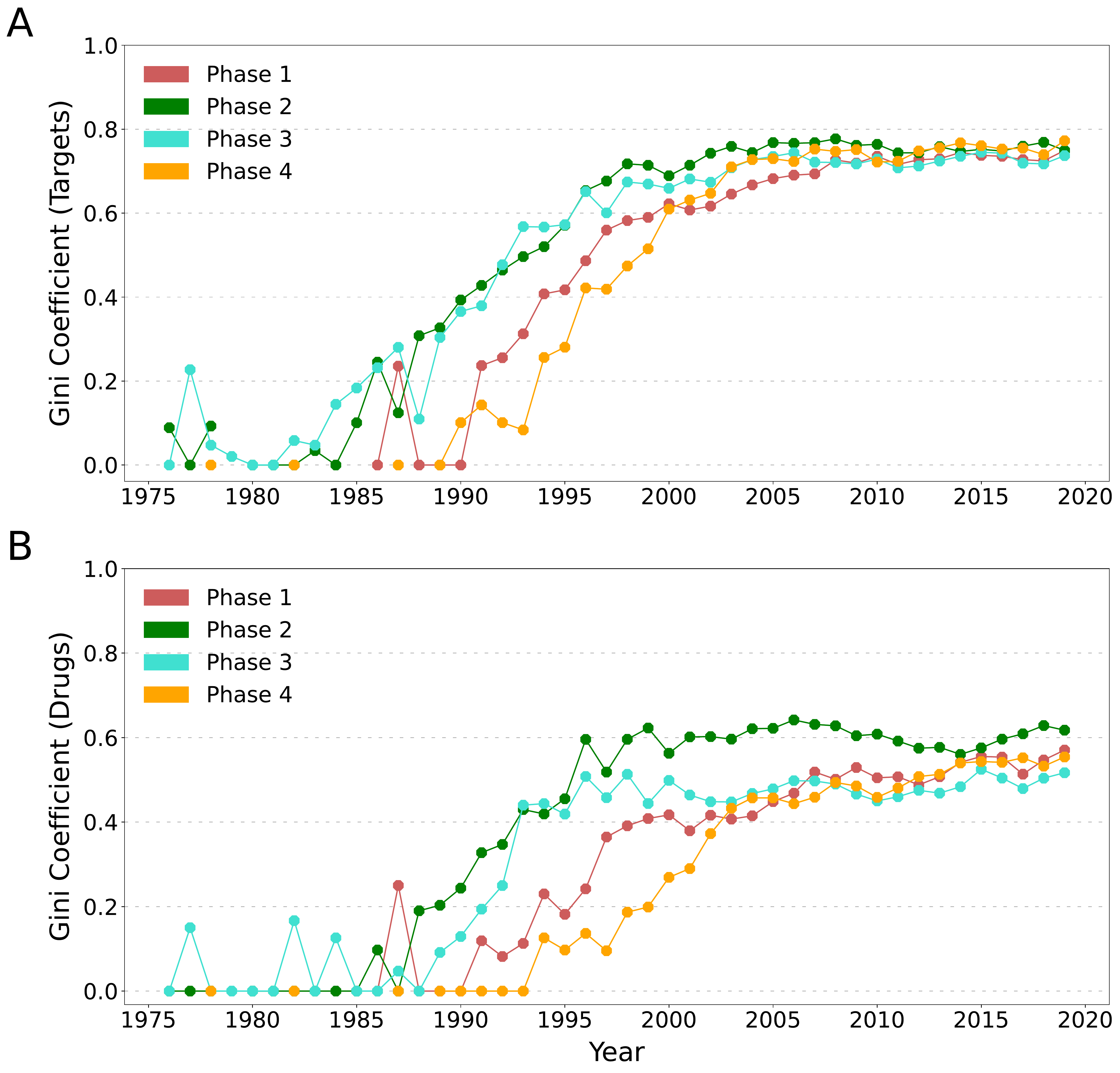}
\caption{\textbf{Inequality in target and drug selection.} (A) Trends in Gini-coefficient targets (B) Trends in Gini-coefficient for drugs. We observe uneven representation of certain targets in multiple trials and over exploration of few drugs in several trials. }
\label{fig:si_gini_targets}
\end{figure}

The lack of novelty in drug trials leads to repeated exploration of previously tested targets. We look at the inequality in target and drug selection using the gini coefficient, where 0 represents complete equality and 1 indicates that all trials test a single target or drug. We find a growing inequality for targets ($Gini_{target} \sim 0.8$) and a growing inequality for drugs ($Gini_{drug} \sim 0.6$) (Supplementary Supplementary Fig \ref{fig:si_gini_targets}), highlighting that a few targets and drugs are tested at high rates. 

\begin{figure}[ht]
\centering
\includegraphics[width=\linewidth]{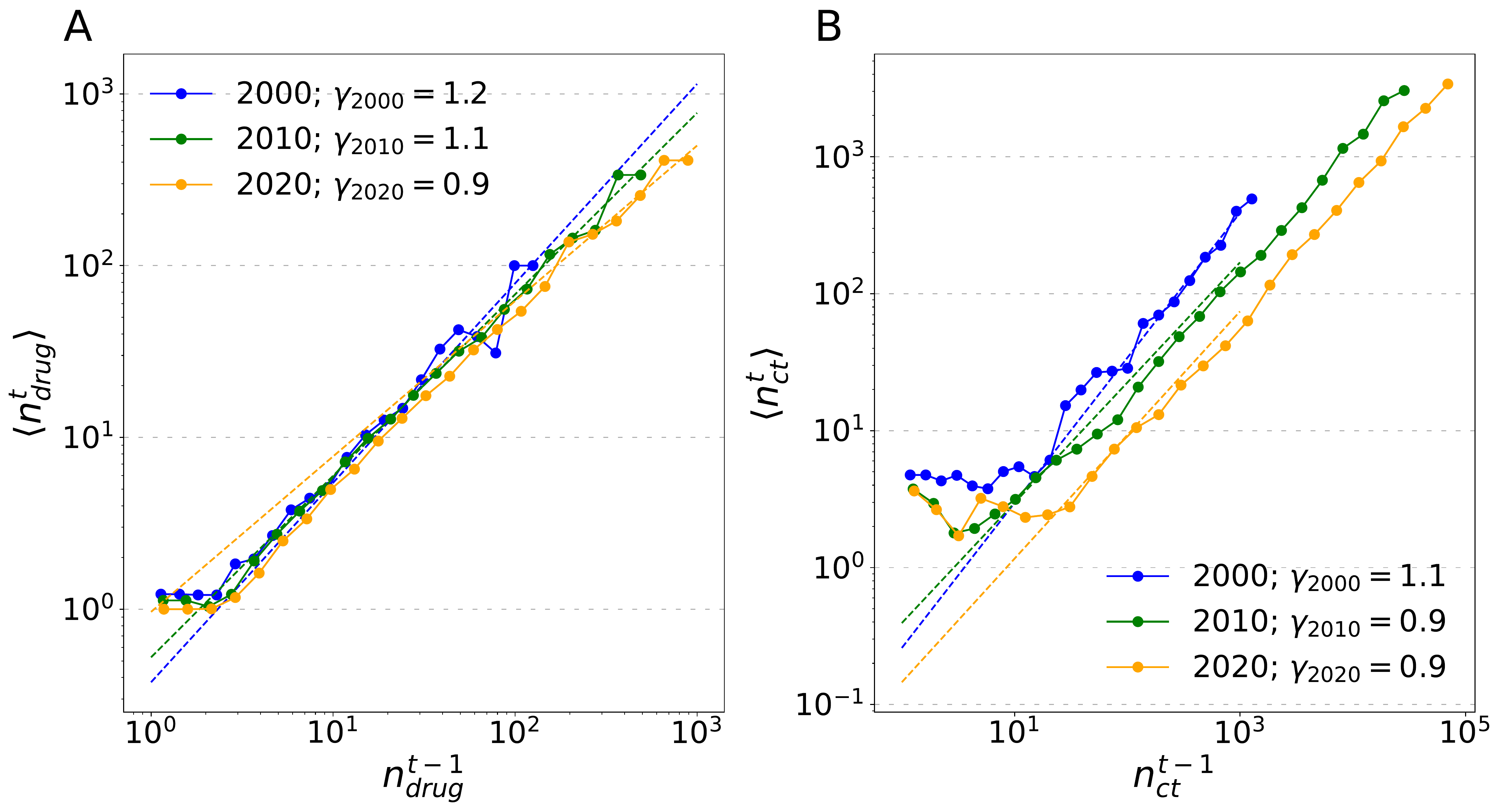}
\caption{\textbf{Evidence of preferential attachment in target exploration.} \textbf{(A)} Number of drugs associated with a target in a given year, $n_{drug}^{t}$ and the number of drugs in the past, $n_{drug}^{t-1}$. \textbf{(B)} Number of clinical trials featuring a target in a given year, $n_{ct}^{t}$ and the number of trials in the past, $n_{ct}^{t-1}$. We observe linearity of the curves for years 2000, 2010, 2020, offering empirical evidence that the likelihood of a selecting a target for a new drug or a new trial is proportional to the number of drugs or trials in the past.}
\label{fig:pa_evidence_targets}
\end{figure}

\section{Impact of approvals and patents}

\begin{figure}[ht]
\centering
\includegraphics[width=30em]{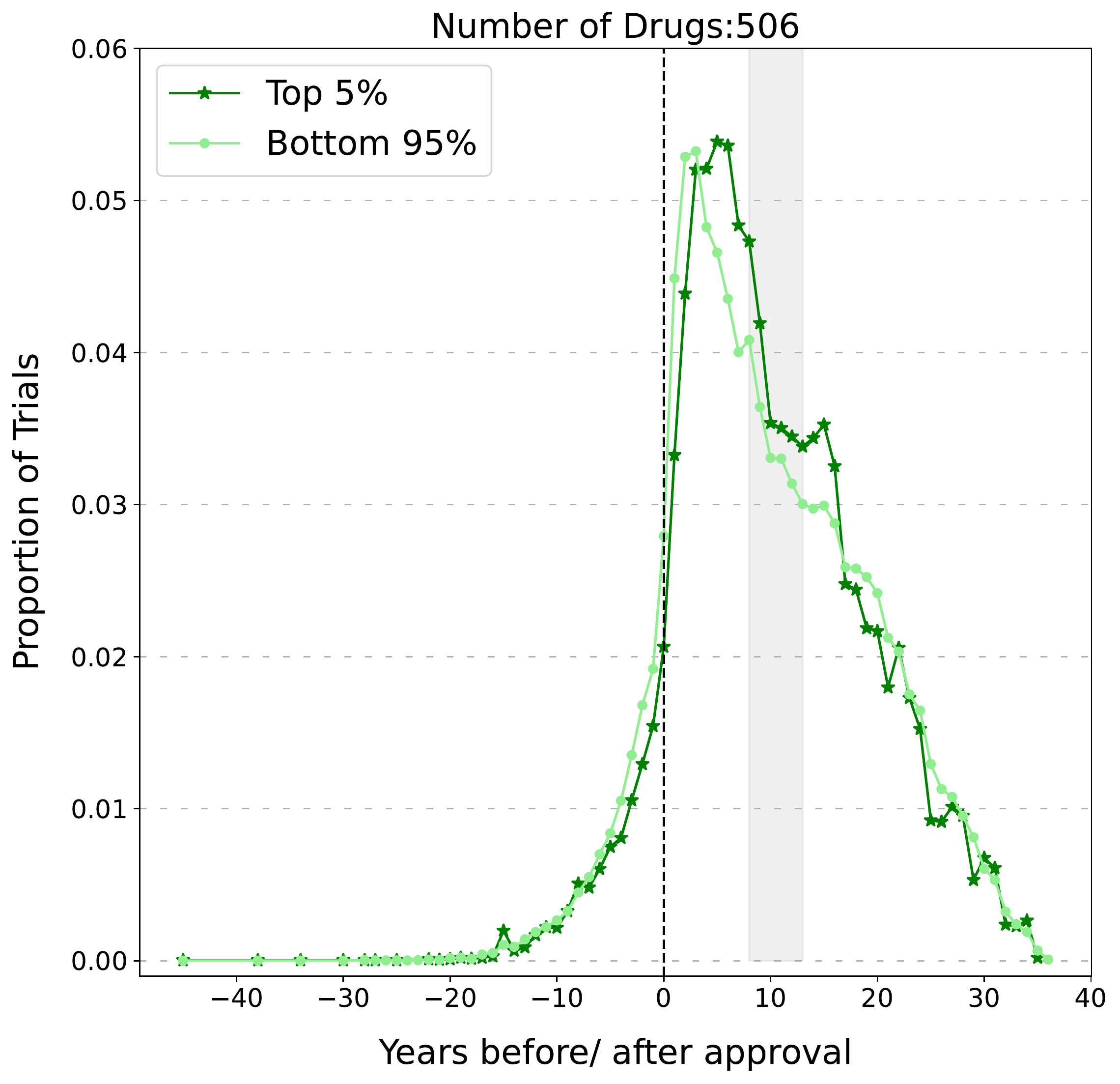}
\caption{\textbf{Impact of patent timeline on trials.} We consider 506 drugs that were granted the approval during our clinical trial data timeline. The patent for a drug lasts from 20 years before the underlying molecule combination become generic and available to other companies for replication. Clinical trials on the drug is expected to take about 8 years prior to receiving approval, estimating anywhere from 8 to 13 years post approval (highlighted) until the drug patent expires. We find increased exploration of approved drugs during this period, following which the drug experiences decreased focus. }
\label{fig:si_approval_num_trials}
\end{figure}

\begin{figure}[ht]
\centering
\includegraphics[width=30em]{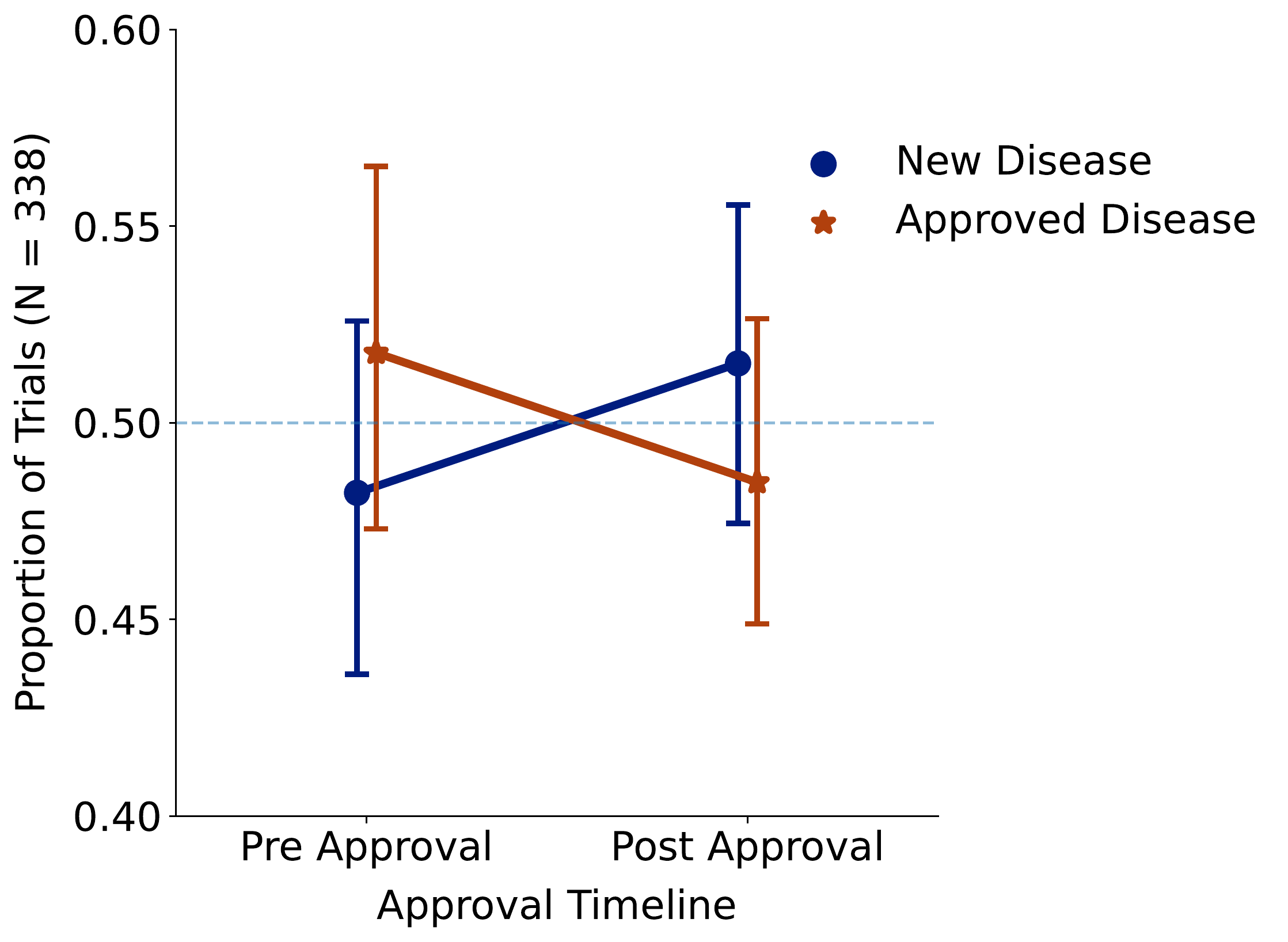}
\caption{\textbf{Impact of approval on repurposing.} We consider all drugs that were granted the approval during our clinical trial data timeline and had a specified approved disease in the FDA database, giving us a list of 338 drugs. We then group trials based on pre and post approval period for the drug and measure the proportion of trials that tested for a disease other than the approved disease. This allows us to measure the impact of approvals in repurposing of the drug for multiple diseases. We find that the trials for drugs post approval test novel diseases than during the pre approval period. The points indicate the average and the lines indicate the 95\% confidence interval. }
\label{fig:si_approval_repurposing}
\end{figure}

Every investigational drug has an estimated patent period for 20 years, after which the exclusive rights for marketing that molecule expires\cite{price2020cost}. We can then estimate $\sim 8$ years until the drug completes clinical trials, providing about 8 to 13 years post approval until the patent expiry. We find that the number of trials for approved drugs increases rapidly post approval (Supplementary Fig \ref{fig:si_approval_num_trials}). Yet, the trials post approval test conditions different from the FDA approved condition, also referred to as drug repurposing (Supplementary Fig \ref{fig:si_approval_repurposing}, indicating that drugs receive increased attention post approval, primarily for diverse diseases. 

\begin{figure}[ht]
\centering
\includegraphics[width=30em]{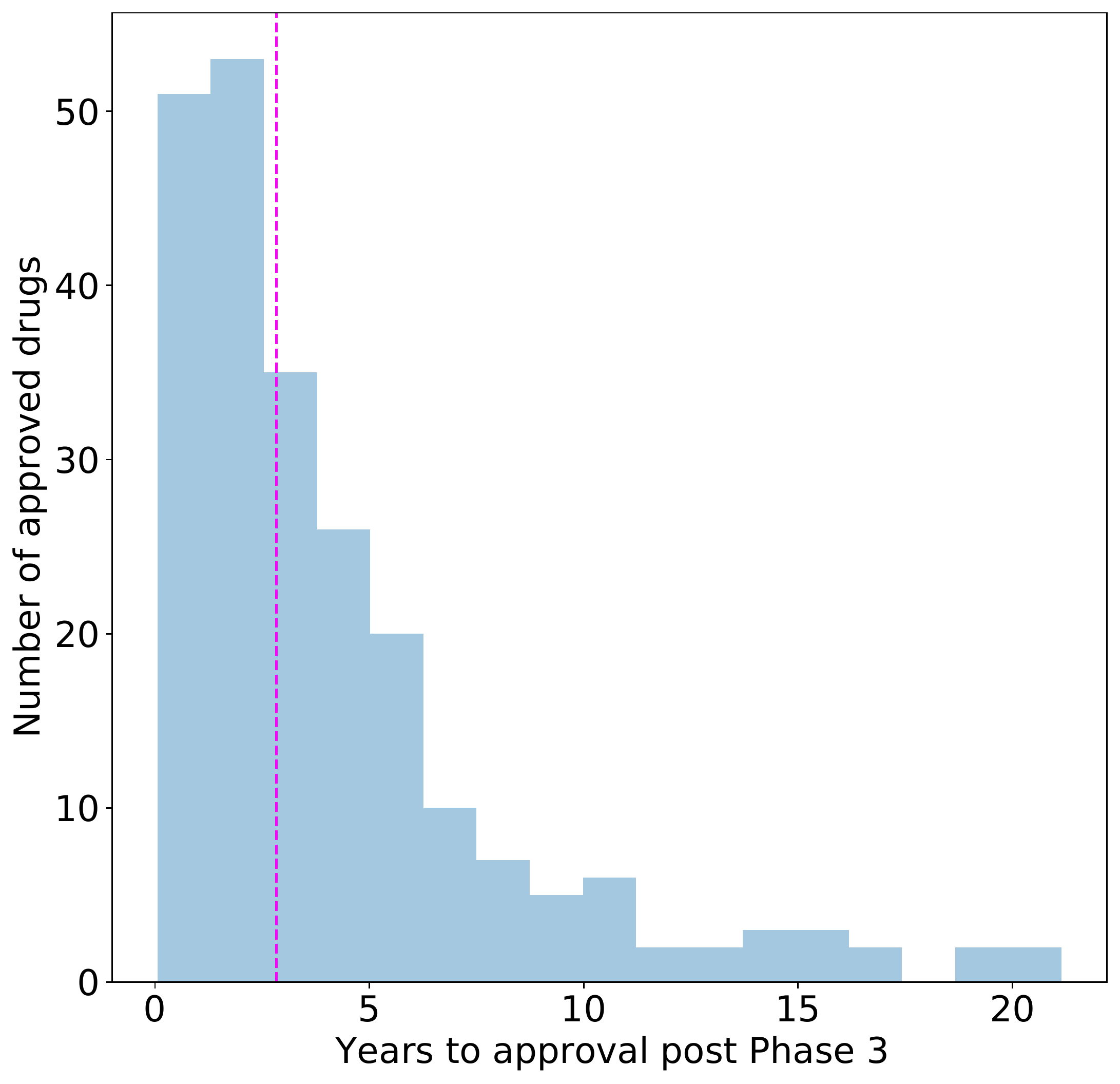}
\caption{\textbf{Time to approval.} We measure the time to approval after clinical development, revealing the time spent on FDA review process. We find that for an average drug it takes 3 years post completion of Phase 3 trial to be approval. }
\label{fig:si_time_to_approval_post_phase_3}
\end{figure}

Finally, we examine the time spent on the FDA approval process. We consider the first completion date of a Phase 3 trial to signal completion of the clinical development. We find that an average drug spends about 3 years after Phase 3 completion to receive approval (Supplementary Fig \ref{fig:si_time_to_approval_post_phase_3}), suggesting a delayed approval period for each drug. 

\section{Mapping the clinical exploration trajectory of proteins}

\begin{figure}[ht]
\centering
\includegraphics[width=30em]{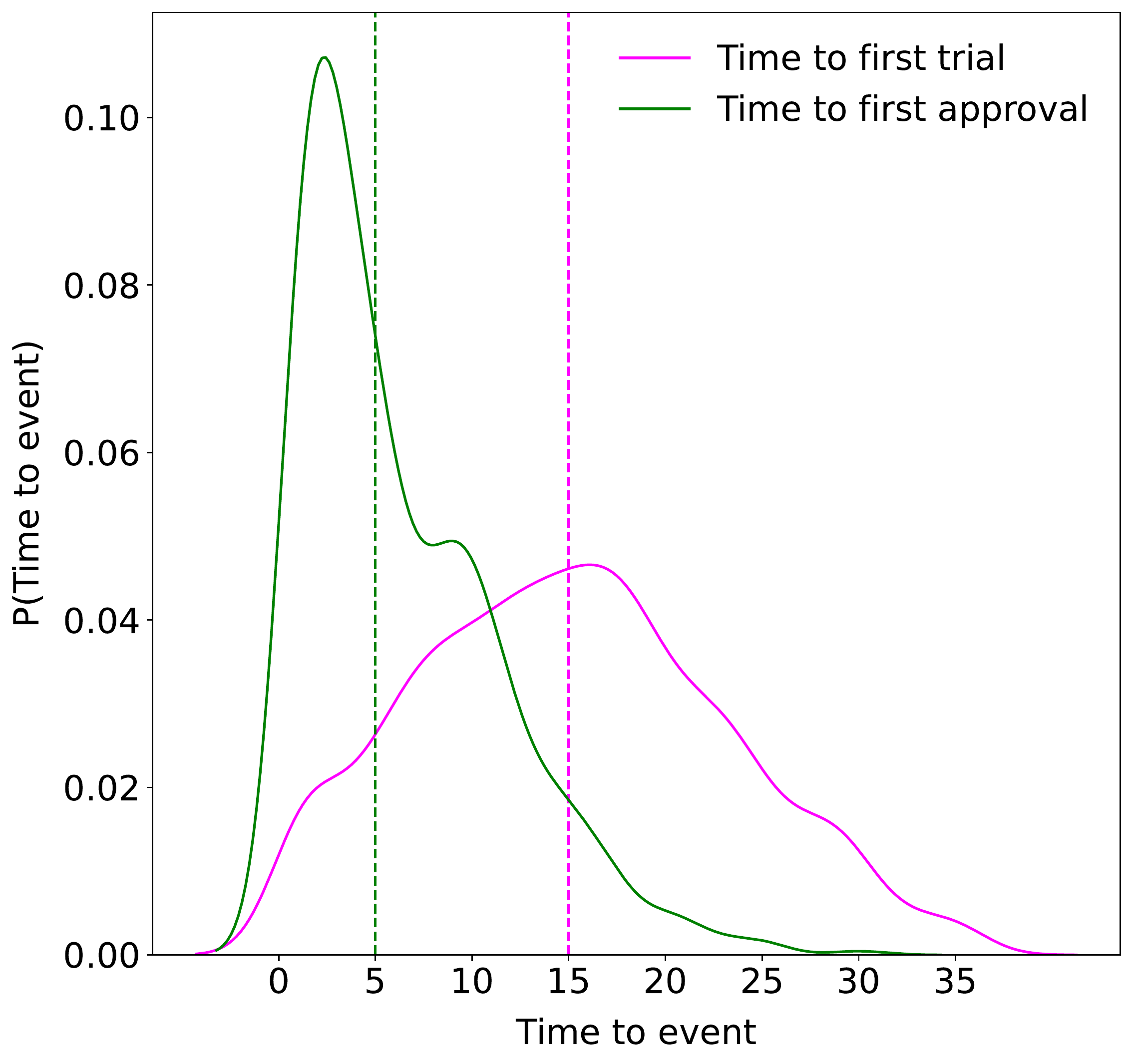}
\caption{\textbf{Time to event.} We measure two variables (i) time to first trial since discovery (magenta) and (ii) time to first approval since first trial (green). The lines indicate the median value for time to event.}
\label{fig:si_time_to_event}
\end{figure}

The first time a protein is associated with a drug in a clinical trial is an important parameter as it represents the year the scientific community recognized the therapeutic utility of that protein. Similarly, the first approved drug of a protein indicates the protein was associated with a drug that showed promising effects in clinical trials. Since new proteins are rarely selected as targets in clinical trials, we next measured the time span between a protein's discovery and its emergence as a target in a clinical trial. We find that it takes a protein, on average, 16 years after its discovery to take part in its first clinical trial and takes another 6 years after the first drug targeting it to receive the first approval (Supplementary Fig \ref{fig:si_time_to_event}). 

\begin{figure}[ht]
\centering
\includegraphics[width=32em, keepaspectratio]{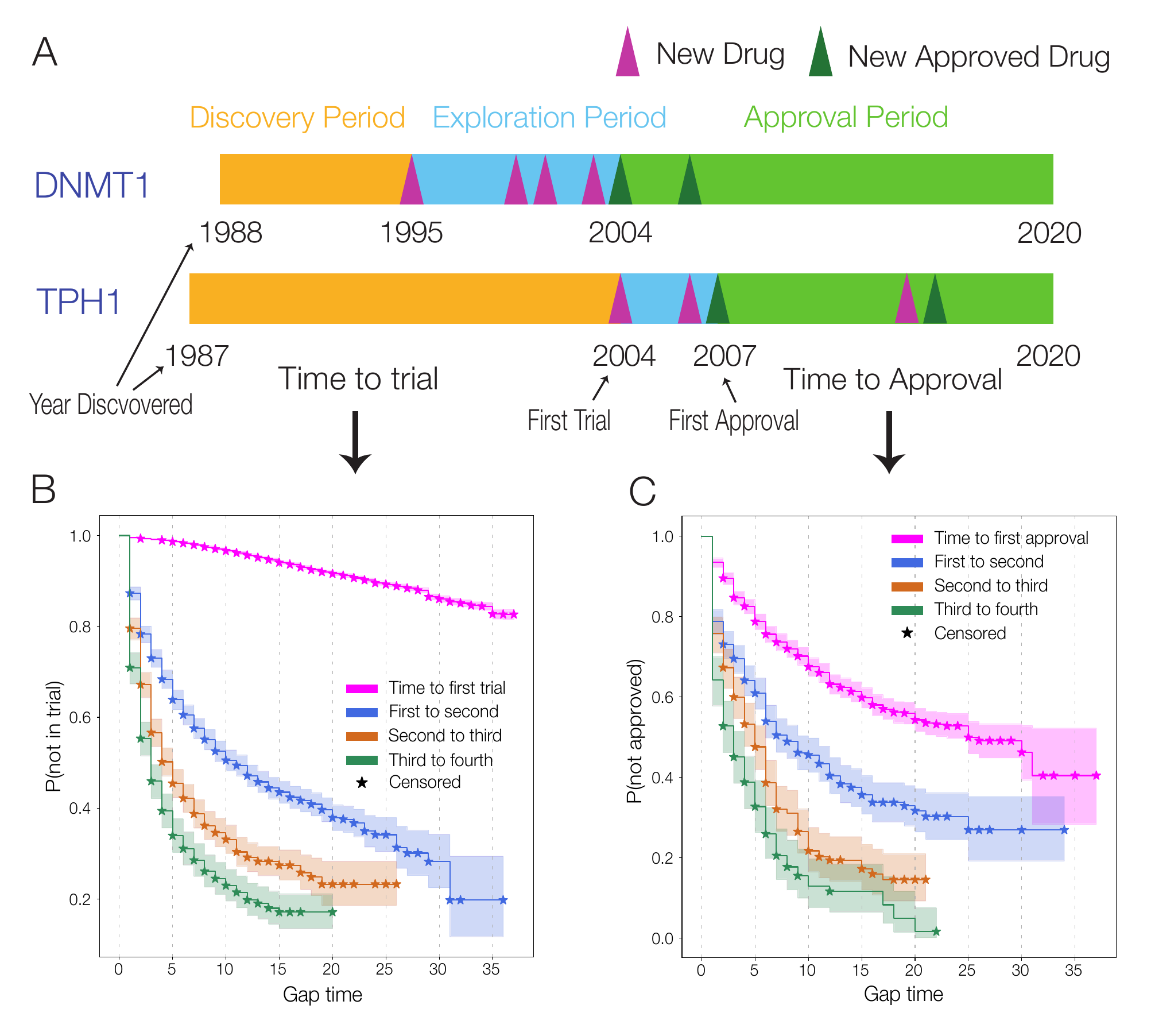}
\caption{\textbf{Repeated exploration of drug-targets.} \textbf{(A)} The exploration trajectory of two targets, DNMT1 and TPH1. We highlight three timeframes;  discovery period (time since discovery to the time of first trial), exploration period (time since first trial to the time since first approval), and approval period (time since first approval to present year). \textbf{(B)} Gap time in exploration, measured as the time spent in discovery period. We observe that the gap time of subsequent new drugs decreases rapidly in that the time to second trial since first trial is lower than the time to first trial, and so on. \textbf{(C)} Gap time in approval, measured as the time spent in exploration period. We find that the time for subsequent approvals decreases as the protein becomes validated as a successful target for multiple drugs. This indicates that targets suffer from visibility bias, hence following the first experimental drug or first approved drug-targeting it, the gap time for subsequent drugs decreases. }
\label{fig:surv_plots}
\end{figure}

For example, the protein DNMT1, which is associated with dementia, bipolar disorder and some rare diseases such as leukemia, pulmonary fibrosis, was discovered in 1988, and first entered clinical trials as a target seven years later, in 1995 (Supplementary Fig \ref{fig:surv_plots} A, top).
Its second trial was in 1999, and the third in 2000.
In 2004, the first drug that targeted the protein was approved, followed by another drug approved in 2006. Similarly, the protein TPH1, which is associated with multiple mental disorders, was discovered in 1987, and its first clinical trial was in 2004, 17 years after its discovery. The second drug was tested in 2006, and the first approved drug emerged in 2007 (Supplementary Fig \ref{fig:surv_plots} A, bottom). 
These exploration patterns prompted us to introduce two variables to quantify recency: 1) time to first trial since discovery of a protein, and 2) time to first approval since the first trial. 

We utilize the Kaplan-Meier survival curves\cite{kaplan1958nonparametric} to estimate the time to event variables. We find that the time to subsequent trials decreases if a protein is targeted by multiple drugs (Supplementary Fig \ref{fig:surv_plots} B), indicating that clinical trials are more likely to focus on recently tested targets. That is, the more drugs target the protein, the more experimental validity it receives, decreasing the time until a subsequent trial. In a similar fashion, the time to approval for targets decreases as it becomes associated with several approved drugs (Supplementary Fig \ref{fig:surv_plots} C), hence the time to second approval is much shorter than the time to first approval, and so on. In summary, we find that proteins experience a long wait time until their first trial as a target, but recently targeted proteins are more likely to be selected for new drugs.  

\begin{figure}[ht]
\centering
\includegraphics[width=40em, keepaspectratio]{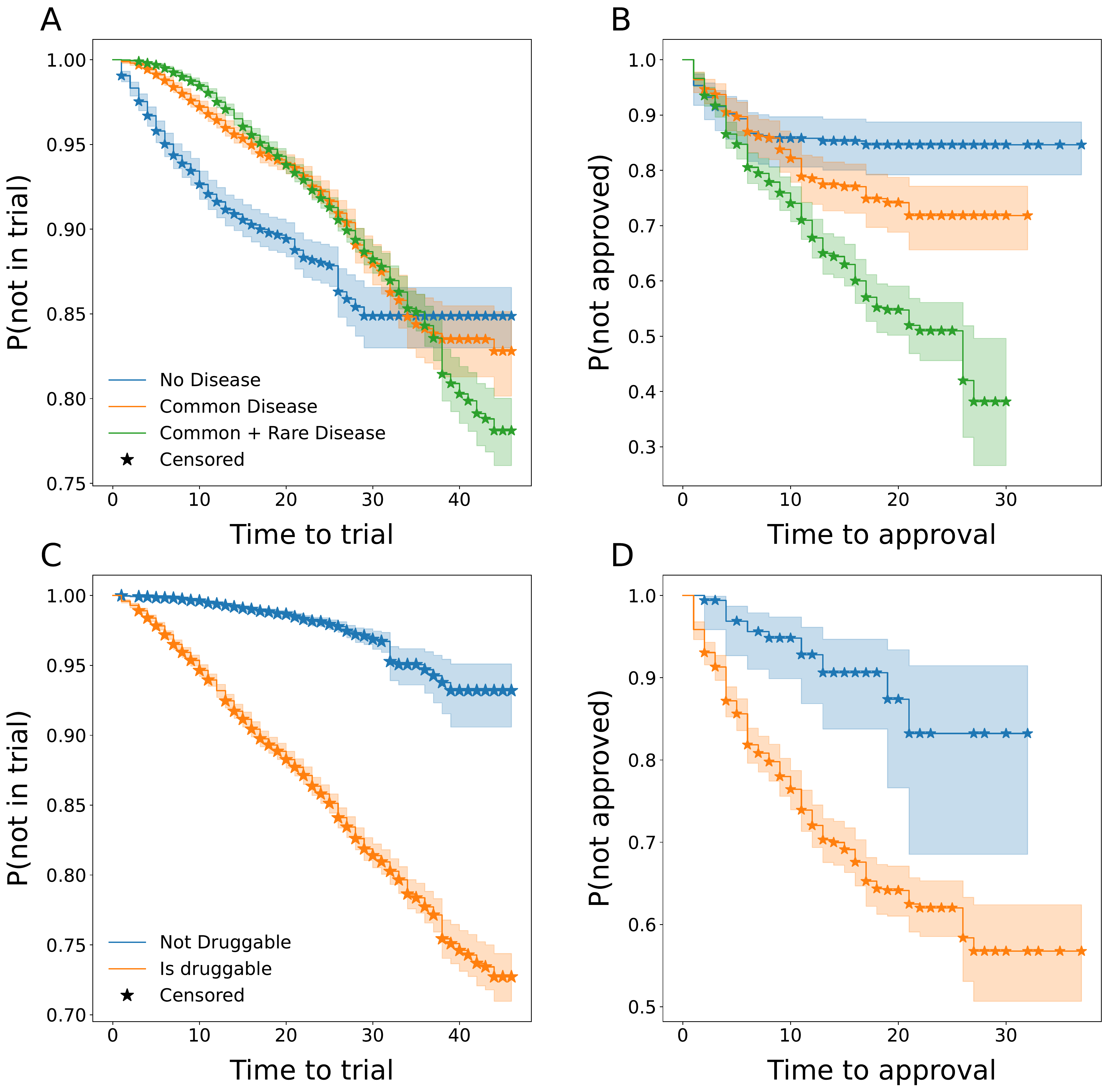}
\caption{\textbf{Survival curves of targets across different classifications.} \textbf{(A)} We classify targets based on its disease association and show the time to first trial since discovery. \textbf{(B)} Time to first approval since first trial for targets based on disease association. \textbf{(C)} Time to first trial since discovery grouped by whether the protein is experimentally verified to be druggable or not. \textbf{(D)} Time to first approval since first trial for targets based on druggability. }
\label{fig:si_surv_plots_classified}
\end{figure}

Further, we find non disease genes enter the trial rapidly after approval but a higher proportion of disease genes eventually receive a trial (Supplementary Fig \ref{fig:si_surv_plots_classified} A). Interestingly, there are no differences in the survival times of common and rare disease genes (log rank test: 0.29, p = 0.58). Further, we find that genes associated with no diseases are less likely to be associated with an approved drug (Supplementary Fig \ref{fig:si_surv_plots_classified} B). Unsurprisingly, druggable genes are more likely to be in a trial and more likely to be approved than non druggable genes (Supplementary Fig \ref{fig:si_surv_plots_classified} C, D). 

\subsection{Repeated occurrence of proteins}

We model the dynamics of repeated occurrences of proteins in trials using the PWP Gap Time model\cite{prentice1981regression}, a survival model for event recurrence estimation, where the time to event resets based on sequential occurrence of events. Specifically, the proteins are stratified based on the clinical trial events, for example, first drug trial, second drug trial. We find that a target's hazard ratio (HR) to be associated to a second drug increases after its first drug trial (HR: 0.82, CI:[0.73, 0.93] vs HR: 1.22, CI:[1.03, 1.45], p<0.01: Supplementary Table \ref{tab:surv_res_trials}), indicating that a protein experiences increased likelihood of a new drug after its first drug trial. In summary, we find that proteins experience a long wait time until their first trial as a target, but recently targeted proteins receive increased attention, reducing the time to be subsequently tested for new drugs.  


\begin{table}[ht]
\centering
\caption{Repeated drug trials of targets}
\begin{tabular}{ l *{6}{c} }
\toprule
\hspace*{2cm} &
\multicolumn{2}{c}{Risk of first trial} &
\multicolumn{2}{c}{Risk of second Trial} &
\multicolumn{2}{c}{Repeated trial} \\
\cmidrule(lr){2-3} \cmidrule(lr){4-5} \cmidrule(lr){6-7}
&
\makebox[3em]{HR} &
\makebox[3em]{95\% CI} &
\makebox[3em]{HR} &
\makebox[3em]{95\% CI} &
\makebox[3em]{HR} &
\makebox[3em]{95\% CI} \\
\midrule
Common Disease & 0.69*** & 0.61, 0.79& 1.23** & 1.02, 1.48 & 0.87 & 0.79, 0.95 \\
Rare and Common Disease & 0.82** & 0.73, 0.93 & 1.22** & 1.03, 1.45 & 0.89** & 0.82, 0.96\\
\midrule
N Total & 18,419& &2,016 & & &  \\
N Censored & 16,403 (89\%)& &956(47\%)&  & & \\
\midrule
\multicolumn{6}{l}{\textsuperscript{***}$p<0.005$,\textsuperscript{**}$p<0.01$, 
  \textsuperscript{*}$p<0.05$}
\end{tabular}
\label{tab:surv_res_trials}
\end{table}

\begin{table}[ht]
\centering
\caption{Repeated approvals of drugs}
\begin{tabular}{ l *{6}{c} }
\toprule
\hspace*{1.5cm} &
\multicolumn{2}{c}{Risk of first approval} &
\multicolumn{2}{c}{Risk of second approval} &
\multicolumn{2}{c}{Repeated approvals} \\
\cmidrule(lr){2-3} \cmidrule(lr){4-5} \cmidrule(lr){6-7}
&
\makebox[3em]{HR} &
\makebox[3em]{95\% CI} &
\makebox[3em]{HR} &
\makebox[3em]{95\% CI} &
\makebox[3em]{HR} &
\makebox[3em]{95\% CI} \\
\midrule
Common Disease & 0.63*** & 0.50, 0.79 & 0.59*** & 0.43, 0.81 & 0.73** & 0.61, 0.87 \\
Rare and Common Disease & 0.75*** & 0.62, 0.91 & 0.73** & 0.57, 0.95 & 0.74*** & 0.64, 0.85 \\
\midrule
N Total & 1,647 & &635 & & &  \\
N Censored & 1,012 (61\%)& &323 (50\%)&  & & \\
\midrule
\multicolumn{6}{l}{\textsuperscript{***}$p<0.005$,\textsuperscript{**}$p<0.01$, 
  \textsuperscript{*}$p<0.05$}
\end{tabular}
\label{tab:surv_res_approv}
\end{table}

\section{Constructing temporal PPI network}

\subsection{Network topology and exploration}

\begin{figure}[ht]
\centering
\includegraphics[width=30em, height=20em, keepaspectratio]{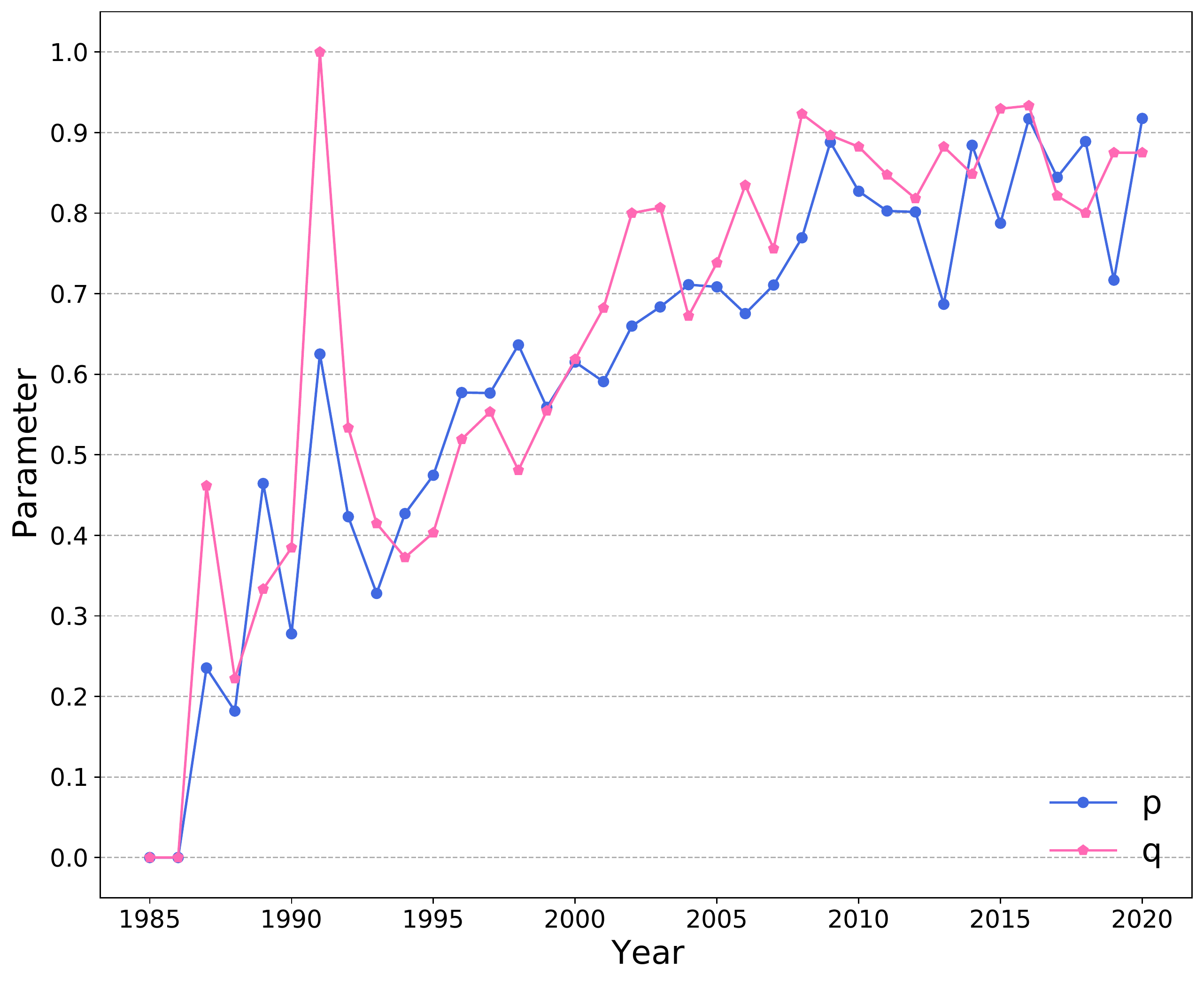}
\caption{\textbf{Characteristics of proteins tested.} We find that genes that are previously tested are preferred at a much higher rate than new targets, represented by parameter $p$ (blue). For untargetted proteins, the proteins part of previously explored neighborhoods are highly preferred, represented by parameter $q$. }
\label{fig:si_net_gene_exploration}
\end{figure}

\begin{figure}[ht]
\centering
\includegraphics[width=\linewidth, keepaspectratio]{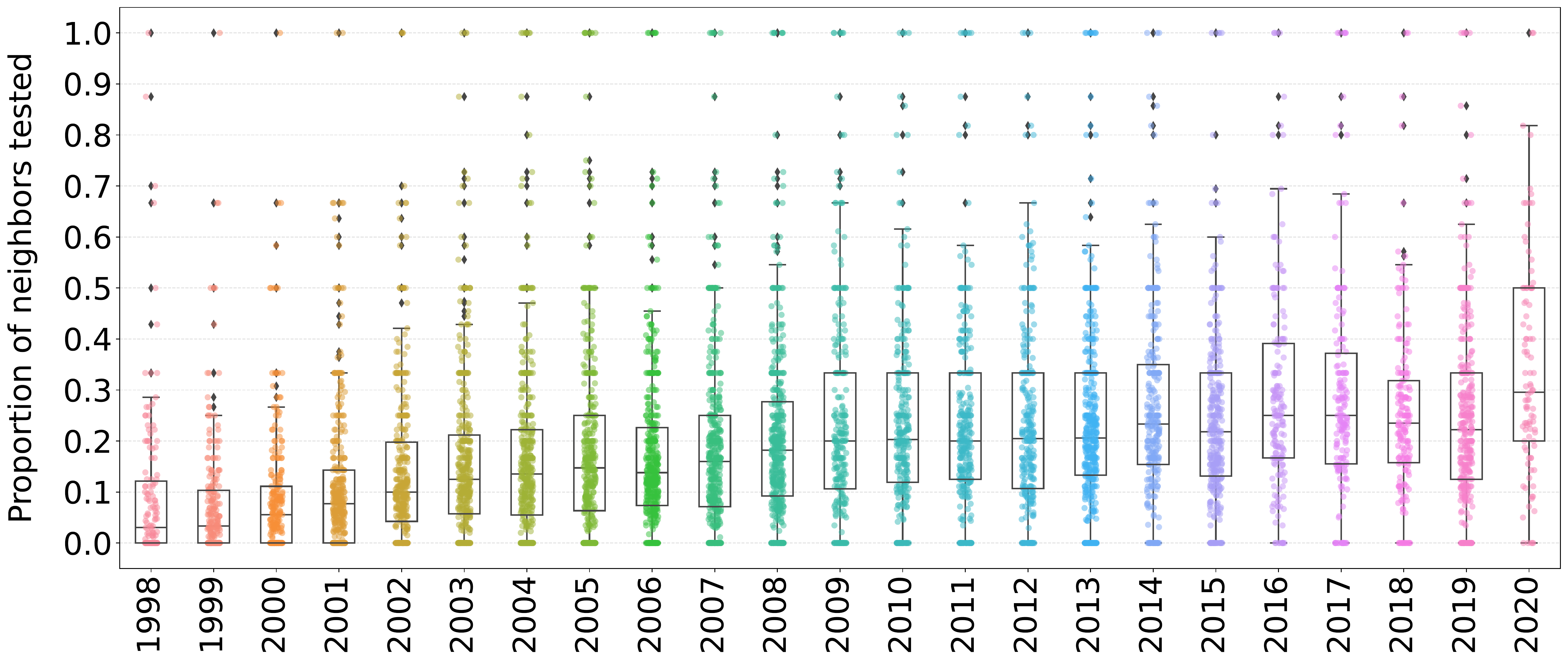}
\caption{\textbf{Local network visibility and exploration.} We find that genes tested in trials have a higher proportion of their network already tested, indicating that the local visibility of the network affects target selection. }
\label{fig:si_prop_net_explored}
\end{figure}

The proteins are connected to other proteins in the well-defined PPI network. We find that proteins embedded in explored neighborhoods are more likely to be selected than proteins with no explored neighbors (Supplementary Fig \ref{fig:si_net_gene_exploration}). Further, proteins with more explored neighborhoods are also preferred. These results highlight that the role of local network structure in new drug exploration (Supplementary Fig \ref{fig:si_prop_net_explored}). These results indicate the network visibility influences exploration of new targets.

\begin{figure}[ht]
\centering
\includegraphics[width=30em, keepaspectratio]{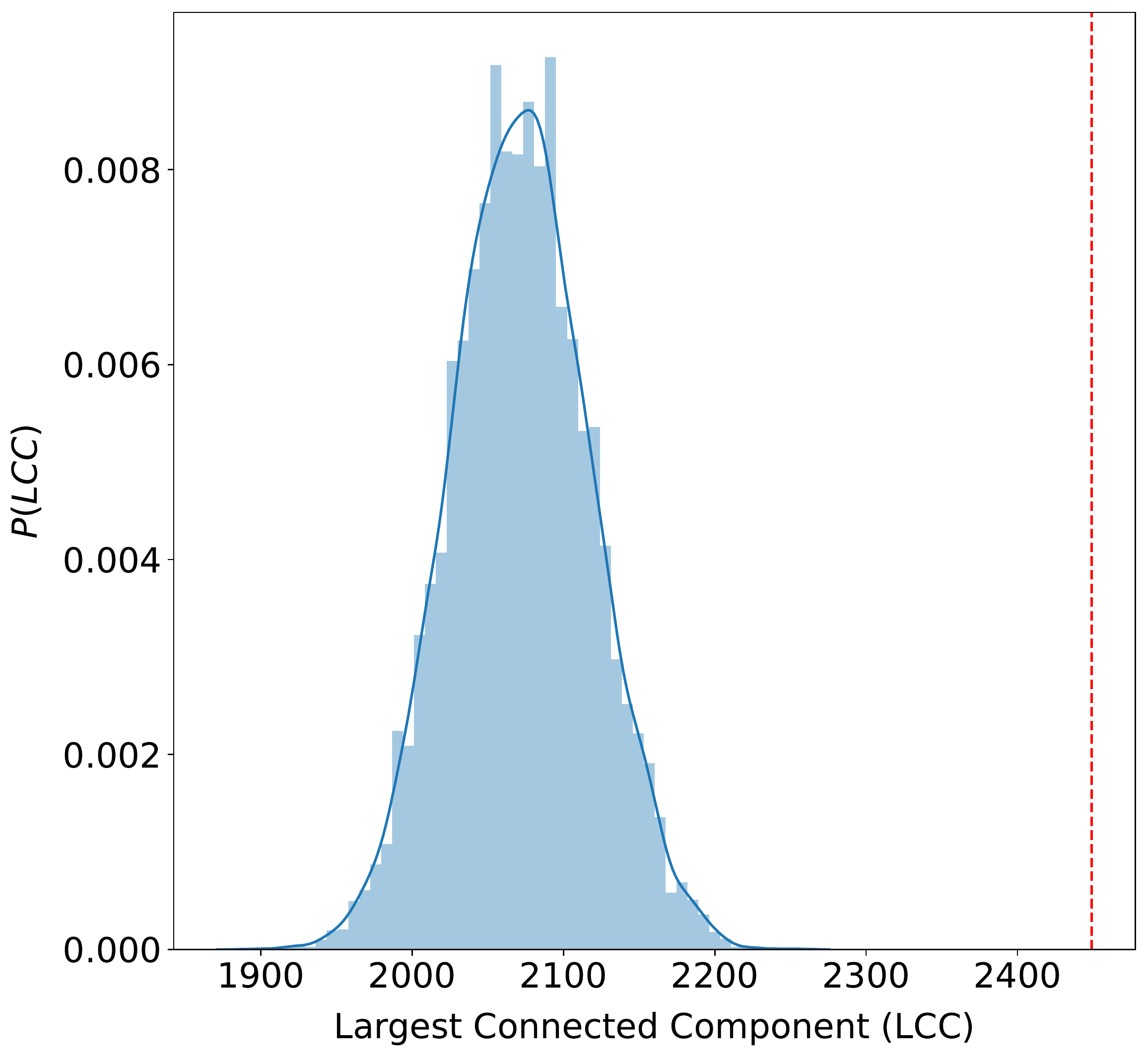}
\caption{\textbf{Largest connected component (LCC) of exploration.} We find a largest component of 2,449 genes (94\%) tested in clinical trials (red line). We random sample 10,000 instances of the same number of genes tested in trials and find that the largest connected component formed is much smaller than found in the empirical network. This indicates that the genes tested in trials are closely connected than expected by random. }
\label{fig:si_lcc}
\end{figure}

\begin{figure}[ht]
\centering
\includegraphics[width=25em, keepaspectratio]{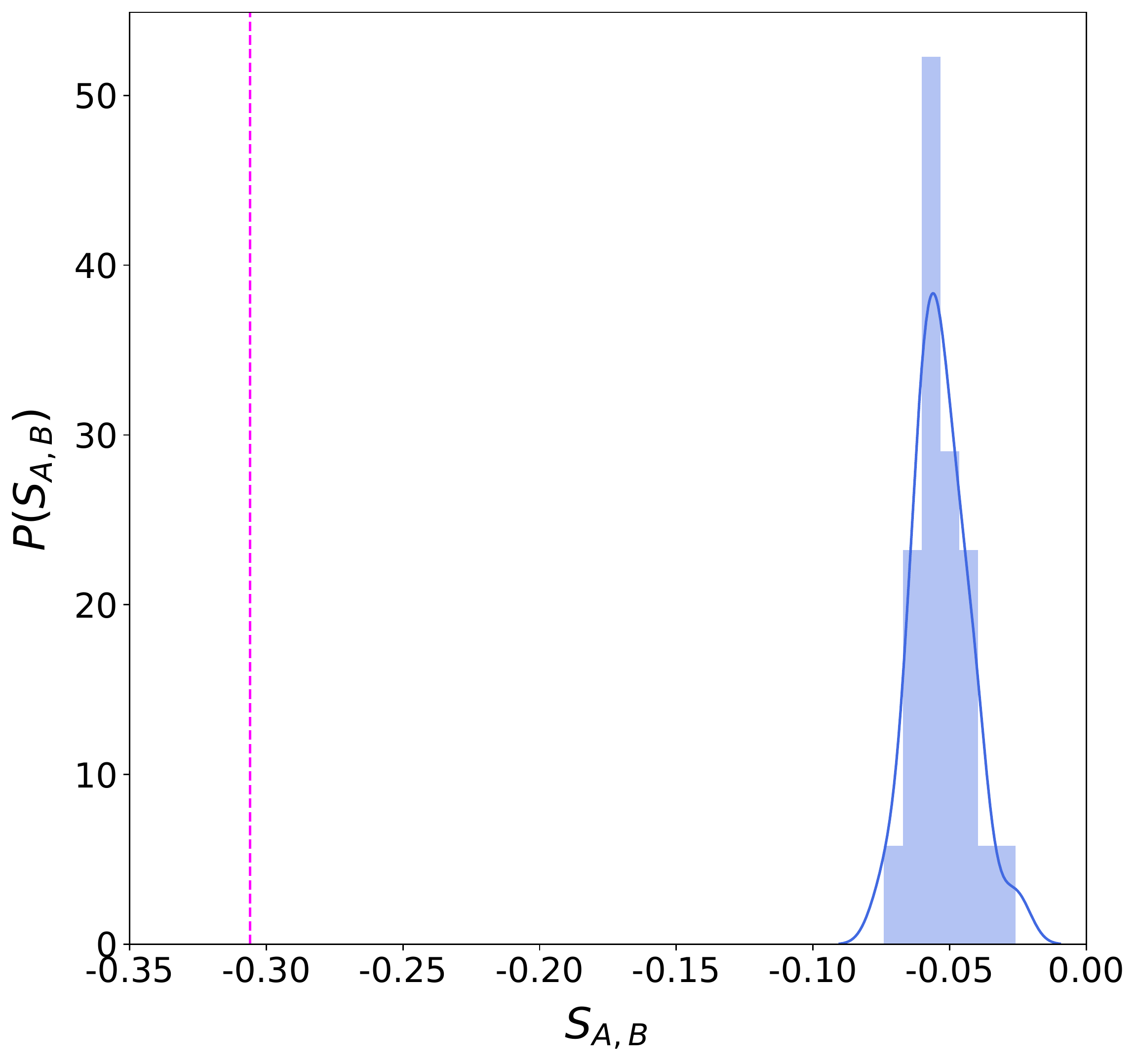}
\caption{\textbf{Separation of proteins that target approved and experimental drugs.} We find that the two groups have overlapping properties than expected at random (blue). Pink indicates the empirical separation. $S_{a,b} = -0.3$, $p <0.001$}
\label{fig:si_sep}
\end{figure}

\textbf{Separation score.} As noted in the main text, 797 (38\%) experimental proteins serve as a target of an approved drug. Of the experimental proteins without target of an approved drug, 891 (76\%) have at least one protein in its local network neighborhood that targets an approved drug, while 274 (23\%) are two degrees away from an approved target. Using the separation score, we test the hypothesis that FDA approval patterns affects drug exploration. We classify proteins into two categories: proteins that are associated with approved drugs and proteins that are associated with experimental drugs. This distinction allows us to measure the separation between the two groups. We use the separation score\cite{menche2015uncovering}, defined using,

\begin{equation}
    S_{A,B} = d_{A,B} - \frac{d_{A,A} + d_{B,B}}{2}
\end{equation}

, where $d_{A,B}$ is the normalized shortest distance between two groups defined as, 
\begin{equation}
    d_{A,B} = \frac{1}{|A|} \sum_{a \in A} \forall_{b\in B} D_{a,b}
\end{equation},

,where $D_{a,b}$, is the shortest distance between two nodes in the network. This formulation allows us to consider $A$ to be the group of experimental proteins and $B$ to be the group of approved proteins. We create random networks, sampling the exact number of proteins found in sets $A$ and $B$, and measure the separation score of the random samples, $S_{A,B}^{r}$ (Supplementary Fig \ref{fig:si_sep}). We then compute the $z_{score}$ using, 

\begin{equation}
    z_{score} = \frac{S_{A,B} - \mu_{S_{A,B}^{r}}}{\sigma_{S_{A,B}^{r}}}
\end{equation}

\begin{figure}[ht]
\centering
\includegraphics[width=25em, keepaspectratio]{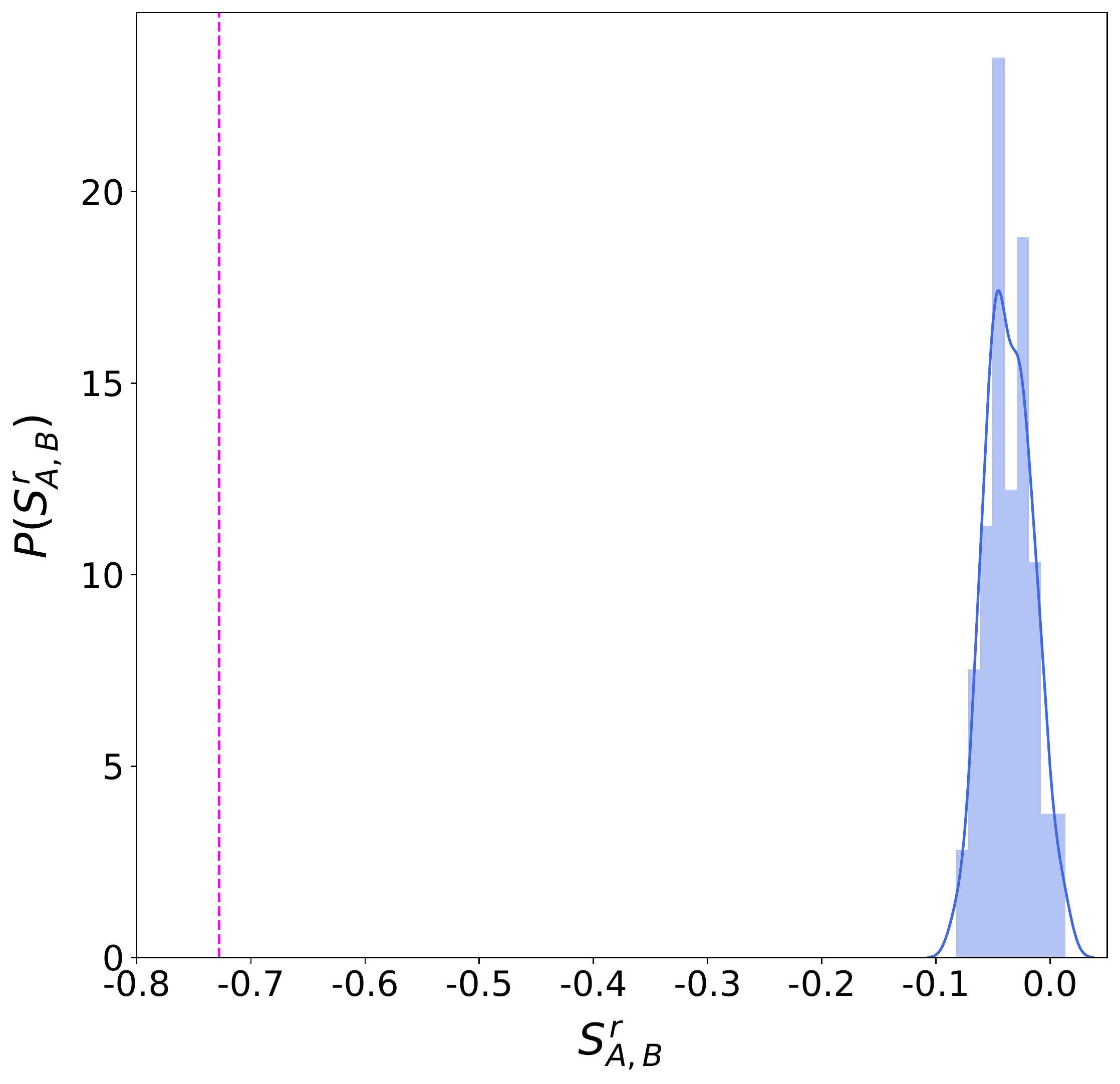}
\caption{\textbf{Separation of proteins associated with experimental and approved drugs.} We utilize the systematically mapped PPI network, comprising of 8,876 proteins and 61,985 interactions. We calculate the separation score between the group of proteins associated with approved drugs and the group of proteins associated with experimental drugs. We find that the two groups are closer in separation compared to the random samples (blue).}
\label{fig:si_sep_hiunion}
\end{figure}

\textbf{Experimentally validated PPI network.} We conduct the same analysis as above for experimentally validated protein interactions, a network comprising of 8,876 proteins and 61,985 interactions. We find the similar result as above, targets of experimental drugs are enriched in the region of proteins that target approved drugs ($p<0.001$; Supplementary Fig \ref{fig:si_sep_hiunion}), verifying that the network processes are not driven by potential selection biases of the PPI network.

\subsection{Defining the inference model}

Scientists and funders have a range of targets mapped to the human genome that could inhibit a disease condition. They collectively choose to pick a select few targets as part of the drug molecule and conduct extensive experiments on that through clinical trials. But what are the underlying processes that guide this exploration?

We create a timeline of PPI network exploration by considering the temporal variable change of multiple protein related parameters. That is, we consider the innovation outlook of the target based on the information available at time $t$, to model the likelihood that a target will be selected at time $t+1$, allowing us to measure the dynamics of network visibility. We use the dependent variables listed below to build a generalized mixed effects model with generalized linear mixed model (GLMM) with target and year as fixed effects. We show the results in Supplementary Table \ref{tab:reg_results} and the results when only considering the primary targets in Supplementary Table \ref{tab:reg_results_primary}). 

\begin{enumerate}
    \item association with a common disease
    \item association with a rare disease
    \item disease associated protein in the neighborhood
    \item number of approved drugs at time $t; n_{approved}^{t}$
    \item number of approved drugs in the neighborhood at time $t; nn_{approved}^{t}$
    \item number of clinical trials at time $t; n_{ct}^{t}$
    \item number of clinical trials in the neighborhood at time $t; nn_{ct}^{t-1}$
    \item number of drugs at time $t; n_{drug}^{t}$
    \item number of drugs in the neighborhood at time $t; nn_{drug}^{t-1}$
\end{enumerate}

\begin{table}[ht]
\caption{\textbf{Logistic model results (Primary and secondary targets.)} Dependent Variable: Target used in trial. Parenthesis indicate standard error.}
\begin{tabular}{l c c c c}
\toprule
 & Model 1 & Model 2 & Model 3 & Model 4\\ 
\midrule
Intercept & -9.68*** & -6.29*** &-6.56*** & -6.06***\\
& (0.24) & (0.22) & (0.24)&(0.23)\\[5pt]
Common Disease & 0.82**   &  0.95*** & 0.66*** & 0.74***\\
& (0.18) & (0.13) & (0.13)&(0.12)\\[5pt]
Rare Disease & 0.54*** & 0.49*** & 0.21**&0.22**\\
& (0.1) & (0.04)&(0.06)&(0.05)\\[5pt]
Disease gene (Neighborhood) & -0.27   & & & \\
& (0.11) & & & \\[5pt]
Prev tested (Neighborhood) & 0.37   & & & \\
& (0.11) & & & \\[5pt]
Disease gene x Prev tested (Neighborhood) & 0.12   & & & \\
& (0.11) & & & \\[5pt]
log2(Number of Approved Drugs) & & 1.32***& &\\
&  & (0.02) & &\\[5pt]
log2(Number of Approved Drugs (Neighborhood))& & 0.017 & &\\
&  & (0.013) & &\\[5pt]
log2(Number of Clinical Trials) & & & 0.39***&\\
&  &  & (0.007)&\\[5pt]
log2(Number of Clinical Trials (Neighborhood)) & & & 0.03***&\\
& &  & (0.006)&\\[5pt]
log2(Number of Drugs Tested) & & &&1.02***\\
& &  & & (0.02)\\[5pt]
log2(Number of Drugs Tested (Neighborhood)) & & & & 0.05*** \\
&  &  & &(0.01)\\[5pt]
\midrule
Year & Incl. & Incl. & Incl. & Incl. \\[5pt]
Gene & Incl. & Incl. & Incl. & Incl.\\[5pt]
\midrule
AIC & 17738 & 16455 & 16246 & 15932\\
\bottomrule
\addlinespace[1ex]
\multicolumn{3}{l}{\textsuperscript{***}$p<0.01$,\textsuperscript{**}$p<0.05$, 
  \textsuperscript{*}$p<0.1$}
\end{tabular}
\label{tab:reg_results}
\end{table}

\begin{table}[ht]
\caption{\textbf{Logistic model results (Primary targets)}. Dependent Variable: Target used in trial. Parenthesis indicate standard error.}
\begin{tabular}{l c c c c}
\toprule
 & Model 1 & Model 2 & Model 3 & Model 4\\ 
\midrule
Intercept & -13.31*** & -10.2*** &-9.46*** & -8.78***\\
& (0.73) & (0.64) & (0.48)&(0.46)\\[5pt]
Common Disease & 1.49**   &  1.74*** & 1.74*** & 1.84***\\
& (0.61) & (0.5) & (0.47)&(0.44)\\[5pt]
Rare Disease & 0.58*** & 0.61*** & 0.56**&0.59**\\
& (0.21) & (0.16)&(0.14)&(0.12)\\[5pt]
Disease gene (Neighborhood) & 0.54   & & & \\
& (0.43) & & & \\[5pt]
Prev tested (Neighborhood) & 0.90**  & & & \\
& (0.46) & & & \\[5pt]
Disease gene x Prev tested (Neighborhood) & -0.69   & & & \\
& (0.47) & & & \\[5pt]
log2(Number of Approved Drugs) & & 2.002***& &\\
&  & (0.1) & &\\[5pt]
log2(Number of Approved Drugs (Neighborhood))& & -0.02 & &\\
&  & (0.03) & &\\[5pt]
log2(Number of Clinical Trials) & & & 0.50***&\\
&  &  & (0.01)&\\[5pt]
log2(Number of Clinical Trials (Neighborhood)) & & & 0.006***&\\
& &  & (0.01)&\\[5pt]
log2(Number of Drugs Tested) & & &&1.42***\\
& &  & & (0.03)\\[5pt]
log2(Number of Drugs Tested (Neighborhood)) & & & & 0.0026 \\
&  &  & &(0.0024)\\[5pt]
\midrule
Year & Incl. & Incl. & Incl. & Incl. \\[5pt]
Gene & Incl. & Incl. & Incl. & Incl.\\[5pt]
\midrule
AIC & 5768 & 5683 & 5667 & 5599\\
\bottomrule
\addlinespace[1ex]
\multicolumn{3}{l}{\textsuperscript{***}$p<0.01$,\textsuperscript{**}$p<0.05$, 
  \textsuperscript{*}$p<0.1$}
\end{tabular}
\label{tab:reg_results_primary}
\end{table}

\section{Network model}

\subsection{Empirical validity}

We model choices in drug discovery using two parameters, first is parameter $p$ that represents the probability of selecting a previously tested protein and second is parameter $q$ that represents the probability of selecting a protein part of a previously explored neighborhood. We utilize the entire search space of $p$ and $q$ to simulate alternative exploration strategies and examine its related benefits for drug discovery. We consider drug exploration from 2011 to 2020 in our simulations, sampling the exact number of proteins tested every year, $m(t)$. 

\begin{figure}[ht]
\centering
\includegraphics[width=\linewidth, keepaspectratio]{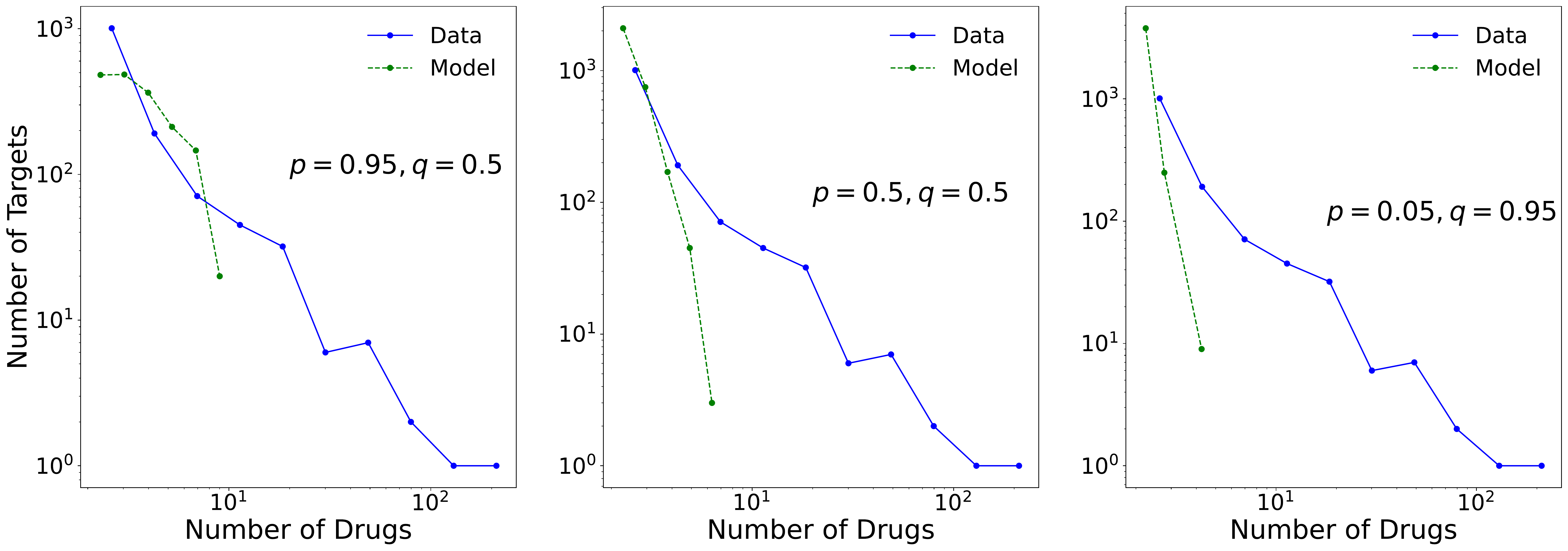}
\caption{\textbf{Absence of preferential attachment.} We simulate the exploration model while removing the preferential attachment aspect of target selection. We display the distribution of number of drugs per target for the parameters \textbf{(A)} $p = 0.95, q = 0.5$ \textbf{(B)} $p = 0.5, q = 0.5$ \textbf{(C)} $p = 0.05, q = 0.95$. We find that the absence of preferential attachment in all three scenarios fails to recreate the empirical distributinos. }
\label{fig:si_pa_absence_model}
\end{figure}

To test the empirical validity of the model, we utilize the resulting distribution of number of drugs per target for each simulation. The distribution characterization how widely proteins are selected as targets for drugs. We utilize the Kolmogorov-Smirnoff distance to measure the maximum difference between the model and the empirical data. As we show in the main text, the model accurately finds this distribution in the preferential attachment (PA) strategy. Yet, we find that the model fails to recreate the observed patterns if we remove preferential selection of drug targets (Supplementary Fig \ref{fig:si_pa_absence_model}).

\textbf{GINI coefficient.} The model selects a list of drugs for each individual target. We measure the statistical dispersion of the distribution of targets using a GINI coefficient, and offers a way to characterize inequality in tested targets. A value of 0 represents complete equality and a value of 1 represents complete inequality. We present the GINI coefficient for the entire search space of the model in Supplementary Fig \ref{fig:si_gini_model}.

\begin{figure}[ht]
\centering
\includegraphics[width=20em, keepaspectratio]{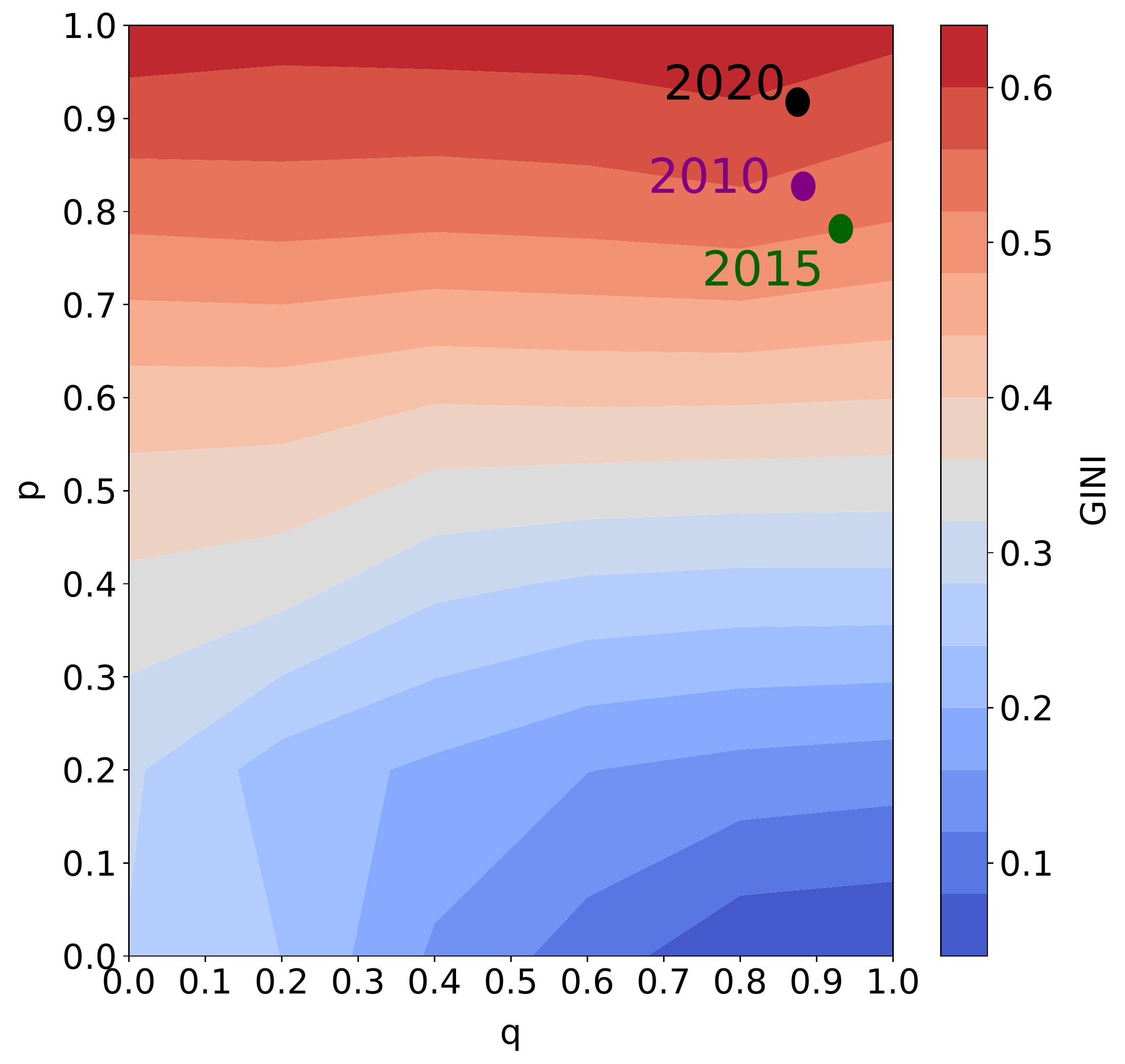}
\caption{\textbf{GINI coefficient.} We measure the inequality in number of drugs per tested targets using a GINI coefficient. 0 indicates complete equality and 1 indicates complete inequality.}
\label{fig:si_gini_model}
\end{figure}

\subsection{Identifying druggable drug targets}

To validate the model's ability to identify potential drug targets, we ask the model to identify drug candidates for three autoimmune diseases - Rheumatoid Arthritis (RA), Crohn Disease (CD), and Asthma. We begin by identifying disease proteins associated to each of the three disease that were tested in previous clinical trials. Next, we search the interaction of these proteins and pick untargeted proteins among them, representing proteins that are part of explored neighborhoods. Next, we use the model to select proteins through the three outlined strategies (PA, R, NS), allowing us to rank proteins based on the frequency they are targeted. Finally, the proteins in the network are validated as druggable, based on extensive experimental studies. We use the well curated list of druggable proteins, \cite{freshour2021integration} to investigate whether the predicted protein has been verified as a potential drug-target, allowing us to measure if the exploration patterns leads to potential druggable outcomes. 

We present the prediction result for the breadth of $p$ and $q$ parameters. Across all three diseases, we find that 70\% of the selected targets through the NS strategy are verified as potential drug candidates (Supplementary Fig \ref{fig:si_prec_cd_asthma}). Indeed, the current practices (PA) selects targets with high accuracy but does so at the cost of prioritizing previously tested targets. In contrast, we show that a network-based search process can be an effective way to improve drug discovery in under-explored regions of the interactome. 

\begin{figure}[ht]
\centering
\includegraphics[width=\linewidth, keepaspectratio]{SI_figures/contour_plot_static_precision_cd_asthma.pdf}
\caption{\textbf{Precision score for the search space.} We utilize the network model to search for new targets and display the precision score for \textbf{(A)} Rheumatoid Arthritis \textbf{(B)} Crohn Disease and \textbf{(C)} Asthma. }
\label{fig:si_prec_cd_asthma}
\end{figure}

\textbf{Target validation.} Additionally, we conduct in-silico studies by searching the predicted results for the network search (NS) strategy. We present the list of identified proteins for RA, CD, and Asthma in Supplementary Table \ref{tab:novel_drug_candidates}, along with the specific functions of each protein, provided by GeneCards\cite{safran2021genecards}.   

The network model is able to find drug candidates in the local network neighborhood of disease-associated proteins. For example, the method selected the protein NLRP3 as a potential drug candidate for RA. NLRP3 interacts with proteins ABCB1, HSP90AA1, CYP3A4, NR1I2, proteins that have been associated to RA and that were previously tested in clinical trials. Indeed, mutations downstream of NLRP3 play an essential role in regulating the inflammasome, identified as a risk factor for inflammatory diseases\cite{villani2009common}. Animal model studies verified that the regulating the over-expression of this gene inhibits the maturation of interleukin-1$\beta$ (IL-1$\beta$), and reduces RA-induced inflammation\cite{liu2020cinnamaldehyde}. These results indicate that the model is able to predict potential novel drug candidates. The illustrated technique can be used to conduct \textit{in-silico} testing of the model predictions for multiple diseases.

\begin{longtable}{|p{5em}| p{2em} | p{5em} | p{5em}| p{18em}|}
\caption{Novel drug candidates discovery}
\label{tab:novel_drug_candidates}
\endfirsthead
\endhead
\toprule
Disease & Rank & Protein & Druggable & Function \\ 
\midrule
Rheumatoid Arthritis & 1 & UBC & True & \small{protein degradation, DNA repair, cell cycle regulation, kinase modification, endocytosis, and regulation of other cell signaling pathways}\\
& 2 & PRKACA & True & \small{transferase activity, transferring phosphorus-containing groups and protein tyrosine kinase activity}\\
& 3 & STUB1 & True & \small{protein homodimerization activity and ligase activity}\\
& 4 & AMFR & False & \small{mediates the polyubiquitination of lysine and cysteine residues on target proteins}\\
& 5 & NLRP3 & True & \small{upstream activator of NF-kappaB signaling}\\
& 6 & GATA2 & True & \small{DNA-binding transcription factor activity and chromatin binding.}\\
& 7 & CTNND2 & False & \small{transcriptional activator and beta-catenin turnover}\\
& 8 & PGRMC1 & False & \small{heme homeostasis, interaction with CYPs}\\
& 9 & KAT2B & True & \small{histone acetyltransferase (HAT) to promote transcriptional activation}\\
& 10 & POR & True & \small{enzyme binding and hydrolase activity}\\
\midrule
Crohn Disease & 1 & UBC & True & {\small protein degradation, DNA repair, cell cycle regulation, kinase modification, endocytosis, and regulation of other cell signaling pathways}\\
& 2 & GK & True & \small{catalyzes the phosphorylation of glycerol by ATP, yielding ADP and glycerol-3-phosphate}\\
& 3 & GATA2 & True & \small{DNA-binding transcription factor activity and chromatin binding.}\\
& 4 & PRKACA & True & \small{transferase activity, transferring phosphorus-containing groups and protein tyrosine kinase activity}\\
& 5 & PGRMC1 & False  & \small{heme homeostasis, interaction with CYPs}\\
& 6 & RAD21 & True  & \small{chromosome segregation, post-replicative DNA repair, embryonic gut development}\\
& 7 & STAT3 & True  & \small{basal beta cell functions, mediates cellular responses to interleukins}\\
& 8 & EP300 & True & \small{histone acetyltransferase that regulates transcription via chromatin remodeling and is important in the processes of cell proliferation and differentiation}\\
& 9 & FOXA1 & True & \small{transcription activator, regulating gene expression in differentiated tissues}\\
& 10 & PRKCD & True & \small{Negatively regulates B cell proliferation, tumor suppressor upon mitogenic stimulation}\\
\midrule
Asthma & 1 & UBC & True  & {\small protein degradation, DNA repair, cell cycle regulation, kinase modification, endocytosis, and regulation of other cell signaling pathways}\\
& 2 & ARVCF & False  & \small{Contributes to the regulation of alternative splicing of pre-mRNAs}\\
& 3 & PKP4 & False  & \small{regulator of Rho activity during cytokinesis}\\
& 4 & GRB2 & False  & \small{link between cell surface growth factor receptors and the Ras signaling pathway}\\
& 5 & PKP2 & False  & \small{transcriptional modulation of beta-integrins}\\
& 6 & KPNA6 & True  & \small{nuclear protein import as an adapter protein for nuclear receptor KPNB1}\\
& 7 & GATA2 & True & \small{DNA-binding transcription factor activity and chromatin binding.}\\
& 8 & YBX1 & True  & \small{numerous cellular processes including regulation of transcription and translation, pre-mRNA splicing, DNA reparation and mRNA packaging}\\
& 9 & PRDM14 & True  & \small{up-regulates the expression of pluripotency gene, proximal enhancer}\\
& 10 & NLRP3 & True  & \small{upstream activator of NF-kappaB signaling}\\
\bottomrule
\end{longtable}


\end{document}